\documentclass[aps,physrev,preprint,superscriptaddress]{revtex4-2}
\usepackage{graphicx} 
\usepackage{amsmath}
\usepackage{subfig}
\usepackage{subcaption}
\usepackage{epsfig,wrapfig}
\usepackage{epstopdf} 

\usepackage{fancyhdr}
\begin{document}

\title{Taming the single-cylinder scattering through time-modulation - The role of the modulation phase}

\author{J. Yan}
\email{jiaruo.yan@iesl.forth.gr}
\affiliation{Institute of Electronic Structure and Laser, Foundation for Research and Technology-Hellas, 70013 Heraklion, Crete, Greece}
\author{I. Katsantonis}
\affiliation {Institute of Electronic Structure and Laser, Foundation for Research and Technology-Hellas, 70013 Heraklion, Crete, Greece}
\affiliation{Department of Materials Science and Technology, University of Crete, 70013 Heraklion, Crete, Greece}
\author{M. Mostafa}
\affiliation{Department of Electronics and Nanoengineering, School
of Electrical Engineering, Aalto University, 02150 Espoo, Finland}
\author{V. Asadchy}
\affiliation{Department of Electronics and Nanoengineering, School
of Electrical Engineering, Aalto University, 02150 Espoo, Finland}
\author{M. Kafesaki}
\email{kafesaki@iesl.forth.gr}
\affiliation {Institute of Electronic Structure and Laser, Foundation for Research and Technology-Hellas, 70013 Heraklion, Crete, Greece}
\affiliation{Department of Materials Science and Technology, University of Crete, 70013 Heraklion, Crete, Greece}

\begin{abstract}
Dielectric particles with time-modulated electromagnetic properties exhibit intriguing scattering phenomena, entailing unique response and possibilities in meta-structures made of such particles. In this work, we investigate the scattering properties of an infinitely-long cylinder with periodically time-modulated permittivity. We demonstrate parametric scattering amplification, depending on the strength of the modulation, as well as controllable angular scattering pattern, ranging from enhanced forward to enhanced backward, or even scattering cancellation. This angular tunability is achieved through appropriate control of the modulation phase, highlighting its critical role in multi-modal time-modulated scattering systems.
\end{abstract}

\maketitle

\section{Introduction}
 Time-varying media provide an additional degree of freedom for tailoring wave–matter interactions, enabling novel effects and functionalities, unachievable in static media\,\cite{ptitcyn2019time, caloz2019spacetime, wu2019serrodyne, mendoncca2002time, pacheco2020temporal, taravati2020space, lyubarov2022amplified, galiffi2022photonics, pacheco2022time,  ptitcyn2023time, engheta2023four}.
The temporal modulation, in the form of temporal interfaces where a dispersionless medium undergoes sudden changes in material properties, has been first analyzed by Morgenthaler in 1958\,\cite{morgenthaler1958velocity}. Morgenthaler showed that the temporal boundary conditions imply conservation of the wavevector, and that temporal reflection and frequency conversion of incident waves arise. 
The expanding interest, though, in the study of time-modulated media has emerged more recently, largely driven by advances in photonic crystals and metamaterials, which have highlighted the vast potential of engineered spatial variations. 
Extending this paradigm to the temporal domain has opened new ways for wave manipulation. Temporal boundaries, in particular, have enabled a variety of applications, including 
temporal aiming\,\cite{pacheco2020temporal}, anti-reflection temporal coatings\,\cite{pacheco2020antireflection}, extension in the bandwidth of absorbers\,\cite{li2020temporal}, as well as broadband amplification of light\,\cite{mostafa2025broadband}.


On the other hand, if the material properties of a medium vary periodically in time, momentum bandgaps are formed\,\cite{asgari2024theory}, in analogy to the frequency bandgaps of conventional photonic crystals with spatial periodicity. Such systems are thus referred to as photonic time crystals (PTCs).
Due to the temporal periodicity, Floquet theorem is applicable to PTC \cite{caloz2019spacetime},  providing a rigorous framework for analyzing and understanding wave propagation in these media. 
Within the momentum bandgaps of PTC, two modes with complex frequencies are supported by Maxwell’s equations. These correspond to two eigenmodes: one that decays and one that grows exponentially in time. Notably, unlike in spatially periodic systems, the amplifying mode is physically allowed, reflecting the non-conservative nature of time-modulated media and their ability to exchange energy with the modulation. 
Recent studies have shown that structural resonances can expand the momentum bandgap relative to that of a homogeneous bulk medium, for the same modulation strength\,\cite{wang2025expanding, canos2026revealing}.
(Note that amplification in PTCs, unlike the conventional parametric amplification, is not essentially a resonance-requiring process\cite{lustig2023photonic}.)

Structural resonances can be implemented in various ways, e.g. using metasurfaces.
In the context of time-modulated Floquet metasurfaces, analytical approaches based on equivalent circuit\,\cite{mostafa2022coherently, simovski2025electromagnetic}, transition matrix approaches\,\cite{garg2022modeling}, and homogenization techniques based on generalized transition sheet conditions or surface susceptibilities\,\cite{chamanara2019simultaneous, iplikcciouglu2025analytical}\, have been developed and validated with numerical results. Applications including frequency conversion \cite{ramaccia2019phase, stefanini2024time, taravati2021pure}, parametric amplification \cite{wen2022unidirectional, ranadive2022kerr}, electromagnetic absorption\,\cite{mostafa2022coherently, hayran2024beyond}, subwavelength imaging\,\cite{simovski2025electromagnetic}, and nonreciprocal responses\,\cite{sounas2017non, chen2019nonreciprocal,wang2020nonreciprocity} have been reported.

A conceptually easy approach to realize resonant Floquet metasurfaces, in particular in the near-IR and optical region, is the involvement of dielectric scatterers supporting Mie resonances.
In this context, scattering particles of spherical\,\cite{stefanou2021light, ptitcyn2023floquet, asadchy2022parametric} and extended shapes\,\cite{stefanou2023light}, have been investigated under periodic temporal modulation of their relative permittivity,
considering cases with and without intrinsic frequency dispersion. Quasi-bound states in continuum (quasi-BIC) resonances hosted in metasurfaces formed by spherical scatterers have been shown to lower the minimum modulation strength required to achieve lasing and coherent absorption\,\cite{globosits2025exceptional}. Reference \cite{garg2025photonic} also discussed quasi-BIC states sustained in time-modulated metasurfaces composed of multilayered spheres, demonstrating their effectiveness in reducing the modulation strength required to realize wide momentum bandgaps.

Despite the extensive body of work on metasurfaces based on spherical scatterers, time-modulated metasurfaces composed of cylindrical Mie resonators remain largely unexplored. Such structures are realizable, for example, employing plasma tubes at microwave frequencies, or semiconducting or transparent conductive oxide wires at higher frequencies. Combining resonant responses with inherent anisotropy, these structures offer a promising platform for accessing additional and currently unexplored wave phenomena. 

In this work, we make a first step in the direction of time-modulated cylinder-based metasurfaces, by investigating their fundamental building block, i.e. an infinitely long cylinder with periodically time-modulated permittivity. 
Although a cylinder of infinite length cannot be fabricated in practice, we expect our theoretical framework to be well applicable to cylinders of lengths that are finite but much longer than the operational wavelengths.
Our study not only demonstrates the emergence of strong parametric resonances in the time-modulated cylinder, but also reveals the critical role of the modulation phase in the cylinder scattering response. We show that by appropriately tuning the modulation phase the scattered field can be controlled in the angular distribution, enabling even transitions from enhanced forward to enhanced backward scattering. This behavior is expected to be a general feature of multi-mode systems, since the modulation phase is shown to govern the interference among the different harmonics arising from time-modulation, in addition to the simultaneously excited Mie resonant modes. Thus, our results propose modulation phase engineering as a key mechanism for controlling wave scattering, opening new avenues for the design of time-varying metasurfaces. 

The rest of the paper is structured as follows: In Section\,\ref{sec:Form}, we establish the theoretical framework for the single-cylinder scattering problem using Floquet–Mie theory, and derive analytical expressions for the scattered fields as well as the corresponding scattering efficiencies. 
Section\,\ref{sec:ParamMie} examines the conditions for parametric resonances, in terms of the modulation frequency and modulation strength; their relationship to the conventional Mie resonances is also discussed. 
Through the interference between the scattered harmonics associated with different multipolar modes, we demonstrate how the far-field radiation can be engineered in Section\,\ref{sec:Rad}. We explore the additional degrees of freedom provided by negative-frequency scattered components, and by the harmonic phase shifts induced by the time-modulation. We show that, by properly selecting the modulation parameters to simultaneously excite multipoles of different orders and by controlling the modulation phase, distinct radiation patterns can be achieved at the same input frequency.

\section{Formalism} \label{sec:Form}
\vspace{-1ex}
We consider a single infinitely long cylinder of material with relative permittivity $\varepsilon_r$ and permeability $\mu_r$, embedded in a homogeneous host medium of relative permittivity $\varepsilon_h$ and permeability $\mu_h$.
Assuming a simple temporal variation of the relative permittivity of the cylinder in the sinusoidal form, and expanding it in Fourier series we have
\vspace{-1ex}
\begin{equation} \label{eq:mod}
    \varepsilon_r(t) = \varepsilon_{ r0}+\Delta\varepsilon\cos(2\pi f_m t+\beta) = \sum_{n=-N}^{N}\varepsilon(n)\exp(in 2\pi f_m t),
\end{equation}
\noindent where $\varepsilon_{ r0}$ is the static permittivity, i.e. the permittivity in the absence of  time-modulation, $\beta$ is the modulation phase, $\Delta\varepsilon$ is the modulation strength, $\varepsilon(n)$ are the Fourier coefficients in the expansion, and $f_m=1/T_{m}$ is the modulation frequency.
To solve the scattering problem for such a time-modulated cylinder, the first step is to find the allowed solutions of the wave equation inside the cylinder.

Applying the Floquet theorem to the electric and magnetic fields (${\bf E}({\bf r},t)$ and ${\bf H}({\bf r},t)$ respectively), and expanding the periodic amplitudes into a Fourier series as in Eq. (1), we obtain ${\bf F}({\bf r},t)=\sum_{n=-N}^{N}{\bf F}^{(n)}(\mathbf{r})\exp{[-i 2\pi f^{(n)} t]}$ (${\bf F}={\bf E}\,\rm{or}\,{\bf H}$), where $f^{(n)}=f-nf_m$ is the $n$-th frequency harmonic supported by the time-modulated medium. Substituting the electric field and the permittivity of Eq. (\ref{eq:mod}) into the wave equation, we obtain
\begin{equation}
    \nabla^2\textbf{E}(t) = \frac{\mu_r}{c^2}\frac{\partial^2\,\varepsilon_r(t)\textbf{E}(t)}
{\partial\,t^2},
\end{equation}
and taking into account that in a homogeneous medium $\nabla^2{\bf E}=-q^2{\bf E}$ with $q$ being the wave number in the medium, we obtain \cite{asgari2024theory,stefanou2021light}
\begin{equation} \label{eq:Eigen_1}
\mu_r(2\pi\frac{f-nf_m}{c})^2 \sum_{p=-N}^{N}\varepsilon(n-p){\bf E}^{(p)}=q^2\mathbf{E}^{(n)},
\end{equation}
where $p$ is an integer.
Eq.\,(\ref{eq:Eigen_1}) is essentially an eigenvalue problem of the form
\begin{equation} \label{eq:eigen}
   \underline{D} \,\underline{M} \mathbf{v}=q^2\mathbf{v},
\end{equation}
where $\underline{D}$ is a diagonal matrix with elements $D_{n',n'}= \mu_r[2\pi(f-nf_m)]^2/c^2$ (with $n'=n+N+1$ to denote the matrix element indices), and $\underline{M}$ is formed by the Fourier coefficients from Eq (1), as $M_{n',p'} = \varepsilon(n-p),$ with $n,p=-N,-N+1,...,N$, $p'=p+N+1$ as matrix element indices.
Solving the problem above, we obtain for a time-modulated infinite medium $2N+1$ eigenvalues, $q_p^2$ (squares of the allowed wavenumbers), $p=-N, ..., N$,  as well as $2N+1$ eigenvectors, $\mathbf{v}_p$ (considered here normalized to unity, with further discussions regarding the normalization in Supplementary B), each containing $2N+1$ elements, $v_{p}(n)$ ($v_{p}(n)$ is the electric field amplitude of the Fourier harmonic with order $n$ contained in the eigenvector $\mathbf{v}_{p}$) 
\cite{stefanou2021light}.

Solving the above eigenvalue problem we can identify the eigenvalues and eigenvectors of the time-modulated medium, and thus express the fields in terms of them. In our case, where the system of interest is a cylinder, it is convenient to express the fields, in and out of the cylinder, in the basis of cylindrical vector harmonics.

Taking the TE polarization as an example, i.e.
${\bf E}$ field perpendicular to the cylinder axis (illustrated as the inset of Fig.\,\ref{fig:Par_Mie}(b)), and considering a 
normally incident (with respect to the cylinder axis) plane wave of  frequency $f$, the field outside the cylinder (sum of the incident and scattered fields) can be expressed as (taking into account that the cylinder will generate fields of frequencies $f^{(n)}=f-nf_m$; see text after Eq. (1)),
\begin{equation}
\begin{split}
    \textbf{E}_{h}(\textbf{r}, t) &= \sum_{n=-N}^N \textbf{E}_{h}^{(n)}({\textbf r})\exp\left[-i2\pi(f-nf_m)t\right]\\
    \textrm{with}\,\,\,\textbf{E}_{h}^{(n)}(\textbf{r}) &= \textbf{E}_{inc}(\textbf{r})\delta_{n0}+\textbf{E}_{sc}^{(n)}(\textbf{r})=i\sum_{l=0}^{\infty} \left[a_l^0\textbf{m}_l(k_0,\textbf{r})\delta_{n0}+a^{(n)'}_{l+}\,\textbf{m}'_l(q_{{h}n},\textbf{r})\right],
    \end{split}
\end{equation}
\noindent where the subscript $h$ represents the host medium, $\textbf{m}_l$ is the $l$-th order vector cylindrical harmonic $\textbf{m}$ (see Supplementary A) expressed in terms of Bessel functions, and $\textbf{m}'_l$ is expressed by replacing the Bessel functions by the corresponding Hankel functions, representing scattered outgoing cylindrical waves. $a_l^0,\,a^{(n)'}_{l+}$ are the expansion coefficients of the incident (0-th Fourier harmonic) and the scattered ($n$-th Fourier harmonic) fields, respectively. Assuming homogeneous host medium, $k_0 =\sqrt{\varepsilon_h\mu_h}\,(2\pi f/c)$ is the wave number of the incident wave, and $q_{{h}n} = \sqrt{\varepsilon_h\mu_h}(2\pi(f-nf_{m})/c)$ is the wave number of the scattered $n$-th Fourier harmonic.
Similarly, the field inside the cylinder can be written as
\begin{equation} \label{eq:E_cyl}
\begin{split}
    \textbf{E}_c(\textbf{r},t) &= \sum_{n=-N}^N \textbf{E}_c^{(n)}(\textbf{r})\exp\left[-i2\pi(f-nf_m)t\right] \\
    \textrm{with}\,\,\,\textbf{E}_c^{(n)}(\textbf{r}) &=i\sum_{p=-N}^N v_{p}(n)\sum_l a'_{l;p}\textbf{m}_l(q_{p},\textbf{r}),
\end{split}
\end{equation}
\noindent where $a'_{l;p}$ is the $l$-th cylindrical harmonic expansion coefficient of the $p$-th eigenmode within the time-modulated cylinder, and $v_p(n)$ is the element of the eigenvector $\textbf{v}_p$. Analogously, we obtain the expressions for the  magnetic field, {\bf H}, as follows: 
 \begin{equation} \label{eq:H_cyl}
\begin{split}
    \textbf{H}_c(\textbf{r},t) &= \sum_{n=-N}^N \textbf{H}_c^{(n)}(\textbf{r})\exp\left[-i2\pi(f-nf_m)t\right]\\
    \textrm{with}\,\,\,\textbf{H}_c^{(n)}(\textbf{r}) &=\sum_{p=-N}^N v_{p}(n)\sum_l b'_{l;p}\textbf{n}_l(q_{p},\textbf{r}).
\end{split}
\end{equation}
(In the above equations $\textbf{n}_l$ is the $l$-th order vector cylindrical harmonic $\textbf{n}$ -  see Supplementary A.)
Applying the boundary conditions equating the tangential components of the electric fields \textbf{E} and magnetic fields \textbf{H}, one can obtain, for each $n$ and $l$, 
\vspace{-1ex} 
\begin{equation} \label{eq:bound_E_TE}
\vspace{-1ex}
    \sum_{p=-N}^{N}a_{l;p}\,v_{p}(n)\frac{\partial{ J_l(q_{p}R)}}{\partial({q_{p}R})}=\frac{\partial{ J_l(k_0R)}}{\partial({k_0R})}\delta_{n0}+a^{(n)}_{l+}\frac{\partial{ H_l(q_{{h}n}R)}}{\partial({q_{{h}n}R})},
\end{equation}
\begin{equation} \label{eq:bound:H_TE}
    \sum_{p=-N}^{N}b_{l;p}\,v_{p}(n)J_l(q_{p}R) = J_l(k_0R)\delta_{n0}+b^{(n)}_{l+}H_l(q_{{h}n}R),
\end{equation}
 \noindent where $R$ is the radius of the cylinder, $J_l, H_l$ are the $l$-th order Bessel function and Hankel function respectively. The scattering coefficients $a_{l;p},b_{l;p}$ and $a^{(n)}_{l+}, b^{(n)}_{l+}$ are all normalized to that of the input wave\,(e.g. $a^{(n)}_{l+}=a^{(n)'}_{l+}/a_l^0$ from Eq. (5)). In addition, we have the relationships: $q_p a_{l;p}=b_{l;p}\mu_r2\pi(f-nf_m)/c$ , $a_l^0 = \sqrt{\mu_h/\varepsilon_h }\,b_l^0$ and $a^{(n)}_{l+}= \sqrt{\mu_h/\varepsilon_h }\,b^{(n)}_{l+}$. {The prime sign denotes coefficients before normalization (which will be discussed later), and the plus sign in the subscript denotes coefficients that represent scattered waves.} (For host medium (e.g. air) with $\varepsilon_h=\mu_h=1$ specifically, the scattering coefficients $a_{l+}^{(n)}=b_{l+}^{(n)}$ are interchangeable for a certain polarization. We solve for $a_{l+}^{(n)}$ in the TE case, and $b_{l+}^{(n)}$ in the TM case and use the corresponding coefficients for later definitions, in order to distinguish the TE and TM cases.)
 Therefore, there are essentially $2(2N+1)$ parameters 
($a_{l;p}$, and $a_{l+}^{(n)}$, $n,p=-N,-N+1\dots N$), numerically solvable based on the above set of simultaneous equations. 
 
The expressions for the TM case can be obtained in a similar way (see Supplementary A), 
and are as follows:
\begin{equation} \label{eq:bound_E_TM}
    \sum_{p=-N}^{N}a_{l;p}\,v_{p}(n)J_l(q_{p}R)
    =J_l(k_0R)\delta_{n0}+a^{(n)}_{l+}H_l(q_{{h}n}R)
    ,
\end{equation}
\begin{equation} \label{eq:bound:H_TM}
    \sum_{p=-N}^{N}b_{l;p}\,v_{p}(n)\frac{\partial{ J_l(q_{p}R)}}{\partial({q_{p}R})}
     = \frac{\partial{ J_l(k_0R)}}{\partial({k_0R})}\delta_{n0}+b^{(n)}_{l+}\frac{\partial{ H_l(q_{{ h}n}R)}}{\partial({q_{{h}n}R})}
     .
\end{equation}

Thereby we obtained the coefficients, for the TE ($a^{(n)}_{l+}$ from Eqs.\,(\ref{eq:bound_E_TE}) and (\ref{eq:bound:H_TE})), or for the TM case ($ b^{(n)}_{l+}$ from Eqs.\,(\ref{eq:bound_E_TM}) and (\ref{eq:bound:H_TM})). 
The total scattering efficiency for the TE mode is obtained as the summation of the contributions from all the harmonics taken into account in the calculations, as (similar to the case of a spherical particle in \cite{stefanou2021light}) 
\begin{equation} \label{eq:time_Qsc}
    Q_{sc}=\sum_{n=-N}^N Q^{(n)}=\sum_{n=-N}^N \frac{2}{\vert q_{{h}n}\vert R }\sum_l \frac{2}{1+\delta_{l0}}\vert a^{(n)}_{l+}\vert^2.
\end{equation}
\noindent {where $Q^{(n)}$ is the scattering efficiency associated with the scattered Floquet harmonic of frequency $f^{(n)}=f-nf_m$.}
For the TM mode, the scattering efficiency is defined by replacing $a^{(n)}_{l+}$ with $b^{(n)}_{l+}$ in Eq.\,(\ref{eq:time_Qsc}).
If $N=0$, the above formalism reduces to the problem of scattering from a time-invariant infinite cylinder \cite{mavidis2020polaritonic}.

\section{Parametric Mie resonances} \label{sec:ParamMie}
{Parametric resonances, where the scattering efficiency is increased significantly, can be realized when the modulation frequency and strength reach certain values. One of the conditions to realize a parametric resonance is to have $f_m=2f$ \cite{canos2026revealing,asadchy2022parametric}. To demonstrate the presence of parametric resonances we consider a dielectric cylinder of radius $R=5$\,mm, $\varepsilon_{r0}=12$. We vary the modulation frequency $f_m$ for a given incident frequency $f=13.35$GHz, and modulation strength $\Delta\varepsilon=1.366$ (which is the calculated threshold modulation strength for parametric resonance of $l=2$ at 13.35\,GHz as we will discuss later). The resulting scattering efficiency for TE polarization is shown in Fig. 1(a). Fig.1(a) demonstrates that, as $f_m$ varies within a wide range, the total scattering efficiency is maximized when the condition $f_m=2f$ is satisfied.}

\begin{figure} 
    \centering
    \subfloat[][]{
    \includegraphics[width=0.4\linewidth]{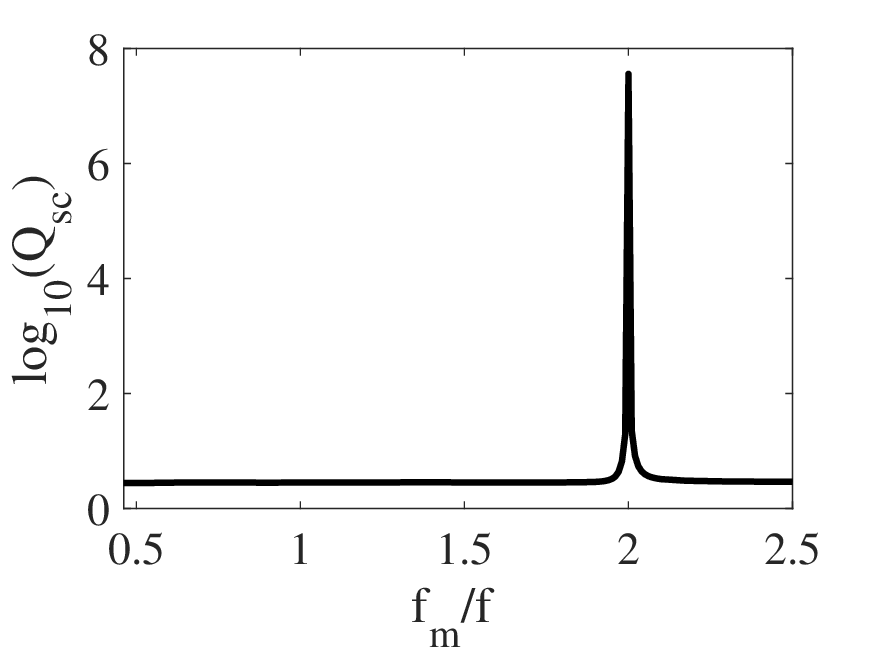}
    }
    \subfloat[][]{
    \includegraphics[width=0.49\linewidth]{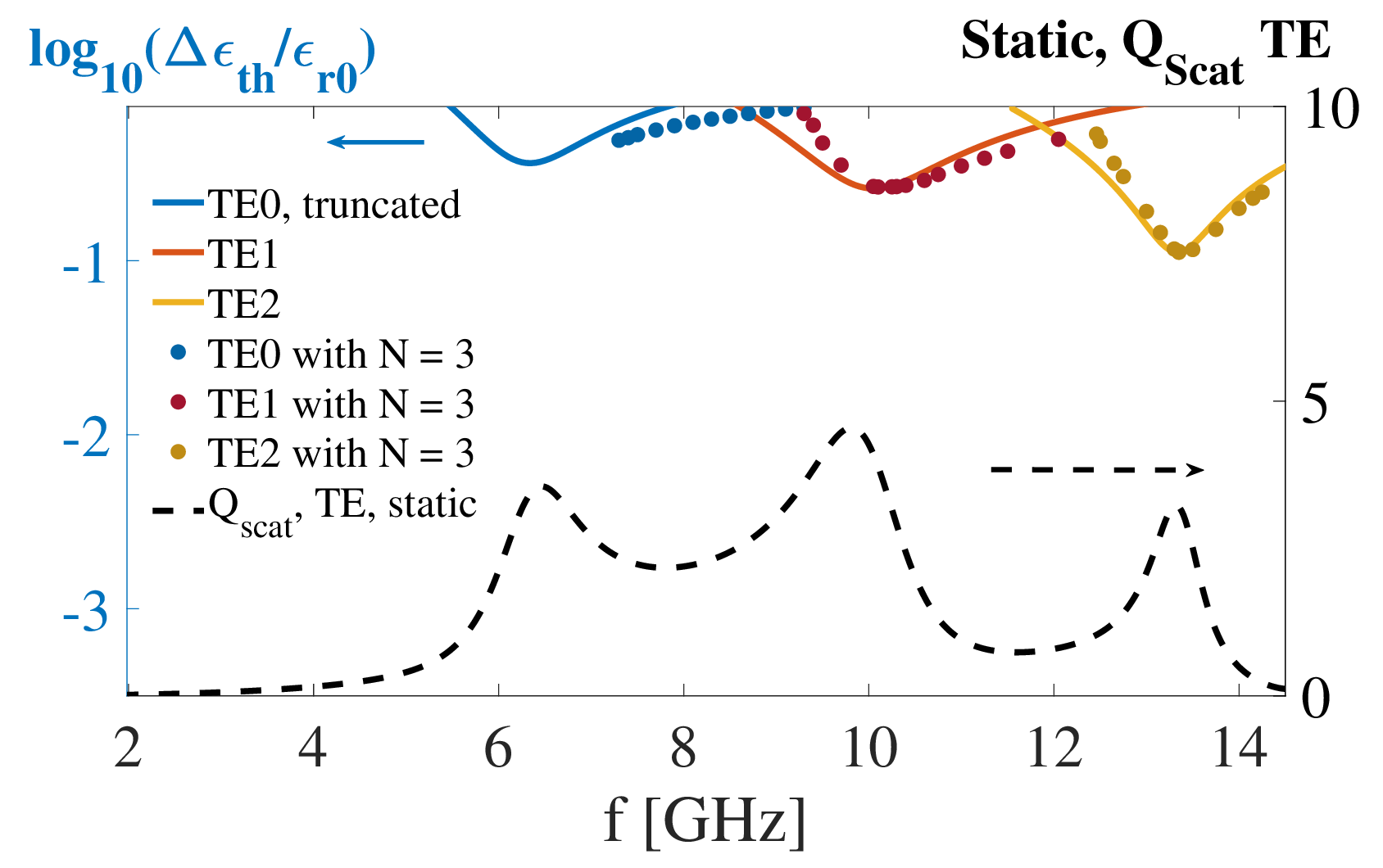}}\hfill
    \subfloat[][]{\vspace{1em}
    \includegraphics[width=0.42\linewidth]{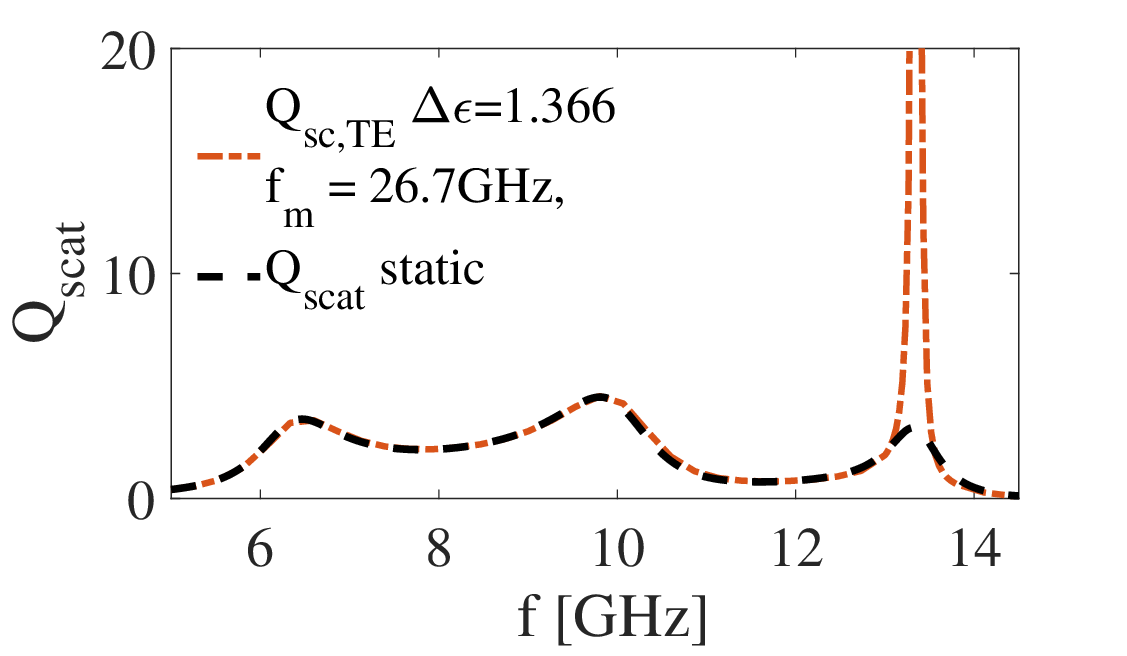}
    }\subfloat[][]{
    \includegraphics[width=0.486\linewidth]{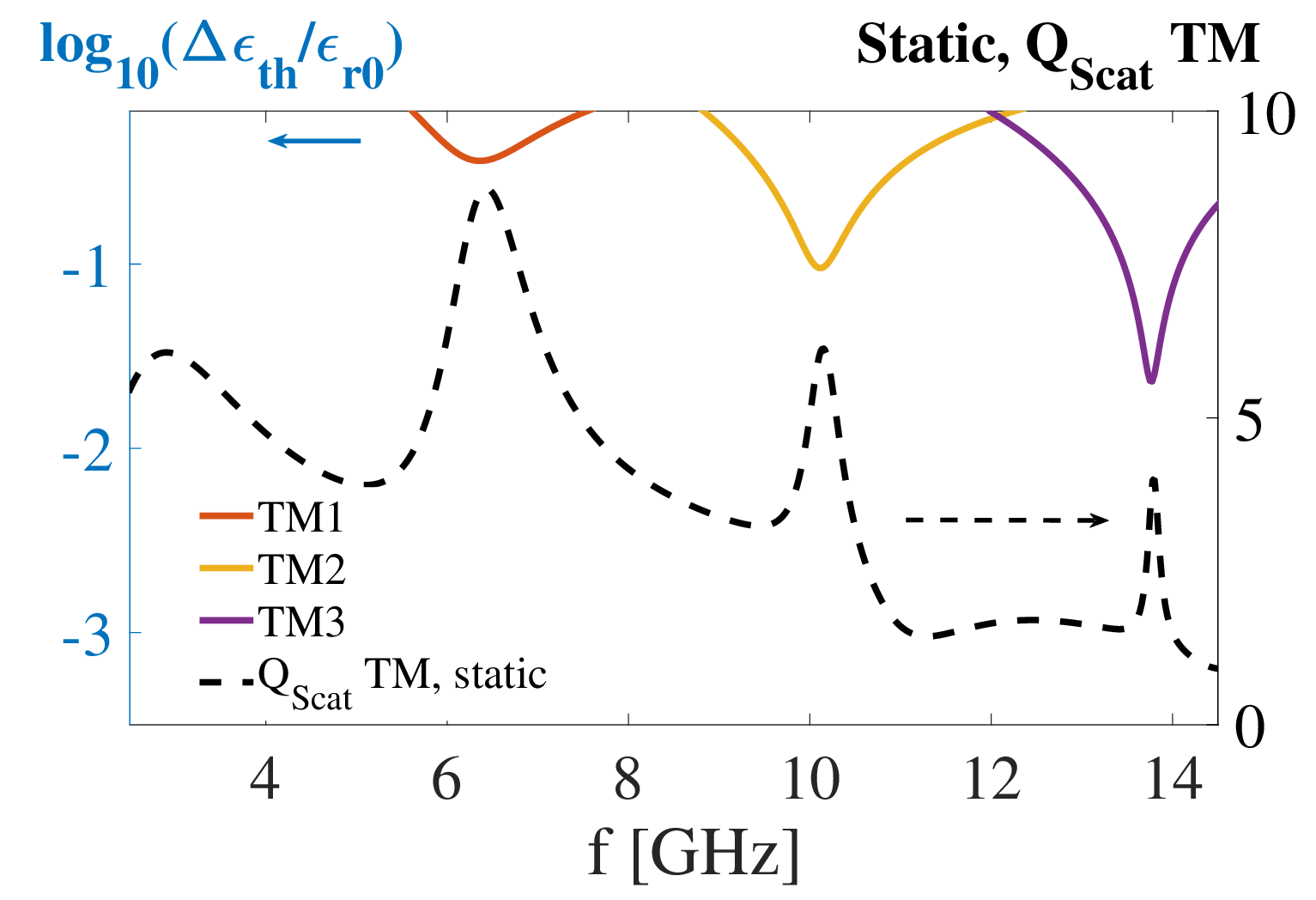}\begin{picture}(0,0)\put(-183,77){\includegraphics[height=1.5cm]{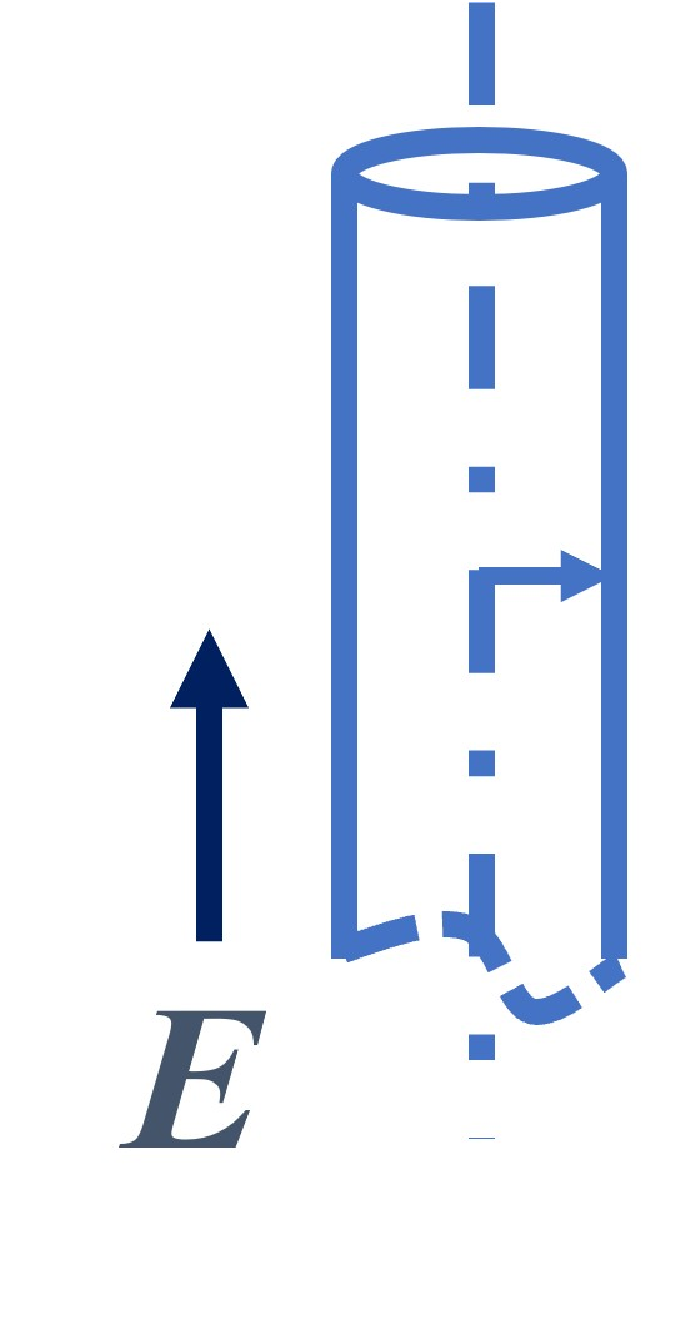}}
\end{picture}}
\caption{{(a) Total scattering efficiency of the TE mode as defined in Eq. (12), evaluated with fixed input frequency $f=13.35$GHz, as the modulation frequency $f_m$ (horizontal axis) is varied. Modulation strength is $\Delta\varepsilon=1.366$. The sharp peak at $f_m/f=2$ indicates the presence of a parametric resonance in this regime. (b) Calculated $\Delta\varepsilon_{th}$ for the TE mode, for each incident frequency $f$ while $f_m$ varies accordingly ($f_m=2f$), referenced to the left axis. Solid curves are based on the truncated eigenvalue problem, Eq.(\ref{eq:th_TE}), and the dots are results taking into account higher-order harmonics. The TE mode $Q_{scat}$ of a static reference cylinder ($\varepsilon_r=12,\,R=5$\,mm) is the black dashed curve, referenced to the left axis. (c) Total scattering efficiency (TE) with $f_m=2\times13.35=26.7$GHz, $\Delta\varepsilon=1.366$, plotted over incident frequency $f$, compared with the unmodulated (static, which is also the black dashed curve in Fig.1(b)) case. } (d) The same as in (b), for the TM case. The minimum threshold modulation strengths for each multipolar mode occur at the Mie resonant frequencies of the static system.}
    \label{fig:Par_Mie}
\end{figure}

To demonstrate the effect of the time modulation on the scattering efficiency for a broad frequency region we first calculate the scattering efficiencies for the static cylinder ($R=5$\,mm, $\varepsilon_r=\varepsilon_{r0}=12$) for both TE and TM polarizations.
The results   are plotted as the black dashed curves in Fig.\,\ref{fig:Par_Mie}{(b) and (d) respectively}, from which we identify the {static} Mie resonances of the system. {The effect of the time modulation on the  total scattering efficiency under $f_m=26.7$GHz and $\Delta\varepsilon=1.366$ is plotted over the incident frequency $f$ in Fig.1(c) for TE waves and compared with the static case. It is clearly seen that significant amplification occurs at 13.35GHz, i.e. the resonance frequency of the static $l=2$, TE mode, which is half of the modulation frequency, satisfying the parametric resonance condition regarding the frequency.}
The conditions for the parametric resonances {regarding the modulation strength} can be established based on Eqs.\,(\ref{eq:bound_E_TE}-\ref{eq:bound:H_TM}) for the eigen-modes, where at the critical modulation  strength the resonances occur without incident fields.

For the cylinder considered in Fig.\,1, first we assume the dominant Fourier harmonics showing parametric resonances (amplification) if the modulation frequency is twice the input frequency, $f_m=2f$, are the $n=0$ and $n=1$ {Fourier} modes. The fundamental scattering mode ($n=0$) is of $f^{(0)}=f$, and the $n=1$ is the corresponding negative-frequency mode $f^{(1)}=-f$, which represents backward propagating waves. {This approximation is justified in the Supplementary, Fig. S2 (b), where the higher-order frequency harmonics are shown to be negligible when parametric resonance is achieved for the TE $l=2$ mode. However, the truncated two-mode approximation is not valid for higher modulation threshold values, as discussed in \cite{asadchy2022parametric} and in more detail in \cite{verde2026optical}. The relevant results, taking into account higher-order harmonics, are discussed later.}

With the modulated relative permittivity taking the form as in Eq.\,(\ref{eq:mod}), the eigenvalues (supported wave vectors within the modulated cylinder) can be obtained as $q_{p=0,1}=(2\pi f/c)\,\sqrt{\varepsilon_{r0}\pm\Delta\varepsilon/2}$ (details can be found in the Supplementary B). The respective eigenvectors are $\textbf{v}_{p=0,1}=\rm[1;\pm e^{i\beta}]$. The subscripts $p=0$ correspond to the plus-sign term in the square-root for consistency.  

The truncated eigen-problem as discussed in \cite{asadchy2022parametric}\,(for spherical scatterers) taking into account only the dominant harmonics ($n=0,\,1$) for the TE mode, can be {formulated based on Eq.(8) and (9), excluding the terms representing the incident waves. The eigen-problem is written as below:}
\begin{equation}
\label{eq:th_TE}
\begin{split}
    \begin{pmatrix}
-\frac{\partial{ J_l(q_{p=1}R)}}{\partial(q_{p=1}R)}\textrm{e}^{i\beta}\hspace{-1em}  & \frac{\partial{ J_l(q_{p=0}R)}}{\partial{(q_{p=0}R)}}\textrm{e}^{i\beta}\hspace{-1em} & -\frac{\partial{ H_l(q_{h1}R)}}{\partial{(q_{h1}R)}} & 0\\ 
\frac{\partial{ J_l(q_{p=1}R)}}{\partial{(q_{p=1}R)}}  & \frac{\partial{ J_l(q_{p=0}R)}}{\partial{(q_{p=0}R)}} & 0 & -\frac{\partial{ H_l(q_{h0}R)}}{\partial{(q_{h0}R)}}\\
-q_{p=1}J_l(q_{p=1}R)\textrm{e}^{i\beta} & q_{p=0}J_l(q_{p=0}R)\textrm{e}^{i\beta} & -\mu_rq_{h1}H_l(q_{h1}R) & 0\\
q_{p=1}J_l(q_{p=1}R) & q_{p=0}J_l(q_{p=0}R) & 0 & -\mu_rq_{h0}H_l(q_{h0}R)
\end{pmatrix} &{\bf .}
\begin{pmatrix}
    a_{l;p=1}\\
    a_{l;p=0}\\
    a_{l+}^{(1)}\\
    a_{l+}^{(0)}
\end{pmatrix}\\
&\,\,\,\,=\textbf{0}
\end{split}
\end{equation}
\noindent where the Bessel and Hankel ($J_l,\,H_l$) functions, and $q_{h0}=2\pi f/c = -q_{h1}$ are as described in Section\,\ref{sec:Form}. $a_{l;p=0}\,{\rm and}\,a_{l;p=1}$ are the wave coefficients within the cylinder, corresponding to the two eigenmodes described by $q_{p=0,1}$ and $\textbf{v}_{ p=0,1}$. Note that the modulation phase induces a phase delay in the scattering coefficient of order $n=1$, i.e. $a_{l+}^{(1)}$. Further discussions regarding the phase induced by time-modulation are given in Section\,\ref{sec:Rad}.
The TM mode eigen-problem then follows as,
\begin{equation}
\label{eq:th_TM}
    \begin{pmatrix}
-\frac{\mu_r\,q_{h1}}{q_{p=1}}J_l(q_{p=1}R)\,e^{i\beta} & \frac{\mu_r\,q_{h1}}{q_{p=0}}J_l(q_{p=0}R)\,e^{i\beta} & -H_l(q_{h1}R) & 0\\
\frac{\mu_r\,q_{h0}}{q_{p=1}}J_l(q_{p=1}R) & \frac{\mu_r\,q_{h0}}{q_{p=0}}J_l(q_{p=0}R) & 0 & -H_l(q_{h0}R)\\
-\frac{\partial{ J_l(q_{{p=1}}R)}}{\partial(q_{{p=1}}R)}\,e^{i\beta}  & \frac{\partial{ J_l(q_{{p=0}}R)}}{\partial{(q_{{p=0}}R)}}\,e^{i\beta} & -\frac{\partial{ H_l(q_{h1}R)}}{\partial{(q_{h1}R)}} & 0\\ 
\frac{\partial{ J_l(q_{{p=1}}R)}}{\partial{(q_{{p=1}}R)}}  & \frac{\partial{ J_l(q_{{p=0}}R)}}{\partial{(q_{{p=0}}R)}} & 0 & -\frac{\partial{ H_l(q_{h0}R)}}{\partial{(q_{h0}R)}}
\end{pmatrix}{\bf .}
\begin{pmatrix}
    b_{l;{p=1}}\\
    b_{l;{p=0}}\\
    b_{l+}^{(1)}\\
    b_{l+}^{(0)}
\end{pmatrix}=\begin{pmatrix}
    0\\
    0\\
    0\\
   0
\end{pmatrix}.
\end{equation}

In order to obtain the eigen-solutions of the above problem, we set the determinant of the above $4\times4$ matrices on the left hand side to zero. This then leads to the approximate conditions {regarding the threshold modulation strength $\Delta\varepsilon_{th}$} for parametric resonances. The eigen-problem is solved for TE and TM modes, {for each frequency}, for each multipolar mode $l$ separately, and the results are plotted in Fig.\,\ref{fig:Par_Mie} (b) and (d).
The horizontal axis is the incident frequency $f$, and the modulation frequency $f_m$ is varied for each $f$ to be $f_m=2f$, as required for achieving parametric conditions. The calculated threshold modulation strengths are referenced to the left axes in log-scale, ${\rm log}_{10}(\Delta\varepsilon_{\it th}/\varepsilon_{\it r0})$. The solid curves of different colors represent results of different multipolar order $l$. {We use the above truncated model only for $\Delta\varepsilon<\varepsilon_{r0}$, so as to ensure $\varepsilon_r(t)>0$ and to avoid the epsilon-near-zero (ENZ) regime where nonlinearity comes into play and the truncated model would fail to capture the underlying physics \cite{verde2026optical}. }

In addition, we show in Fig.\,\ref{fig:Par_Mie}\,(b) (dots) the results of $\Delta\varepsilon_{th}$ taking into account more frequency harmonics. Here the threshold modulation strengths were determined by maximizing the $Q_{{sc}}$ (as defined in Eq.\,(\ref{eq:time_Qsc})), taking into account more harmonics ($N=3$, seven harmonics in total). While truncated eigen-problem is a good approximation for weaker modulations, in cases of stronger modulations, higher-order harmonics are non-negligible. Differences between the results obtained by the two methods (dots and solid curves of the corresponding color) can be seen in Fig.\,\ref{fig:Par_Mie}\,(b), especially for cases where larger $\Delta\varepsilon_{th}$ values are required.
Similar deviations of the threshold modulation strengths between the truncated eigen-problem and the problem involving larger number of harmonics  were discussed also in the supplementary of \cite{asadchy2022parametric}, where 24 harmonics were taken into account. 

The minima of the threshold modulation strengths for each mode occur at $f$ (with $f_m=2f$) close to the corresponding Mie resonances of an unmodulated cylinder of $\varepsilon_{r0}$, demonstrated by the alignments between the dips of the solid curves and the peaks of the dashed curves in Fig.\,\ref{fig:Par_Mie} (b) and (d). Moreover, as discussed in detail in Supplementary C, the minimum values of the threshold modulation strengths are dependent on the quality factors of the respective Mie resonances for the static system. {These effects show that, at least for modulations up to the threshold modulation strength, the modes that dominate the response of the time-modulated system remain the ones of the static system. The dominance of the static-system modes is also verified in \cite{galanis2026perturbative}, where it is further shown that under relatively weak modulation the strengths of the scattering amplitudes of the time-modulated system are dependent on the spectral overlap integrals between the static modes at the incident and scattering frequencies, which can be controlled through geometric tuning or through the time modulation.}

{We have to note here that at the threshold modulation strength, $\Delta \varepsilon_{th}$,  the complex eigenfrequencies of the static cylinder (representing leaky Mie resonances) are being pulled onto the real axis by the parametric gain offered by the time modulation. This is similar to the pump strength of a degenerate parametric oscillator, where the threshold modulation strength that we calculated plays the role of the critical pump amplitude that takes the oscillator from the stable oscillations regime to the parametric instability zone.}
 It should also be noted here that the scattering properties of the unmodulated case are scalable over frequency, dependent on the electrical size of the scatterer. The frequency scalability also applies to the eigen-solutions for obtaining the threshold modulation strengths.
\begin{figure}
    \centering
    \subfloat[][]{
     \includegraphics[width=0.6\linewidth]{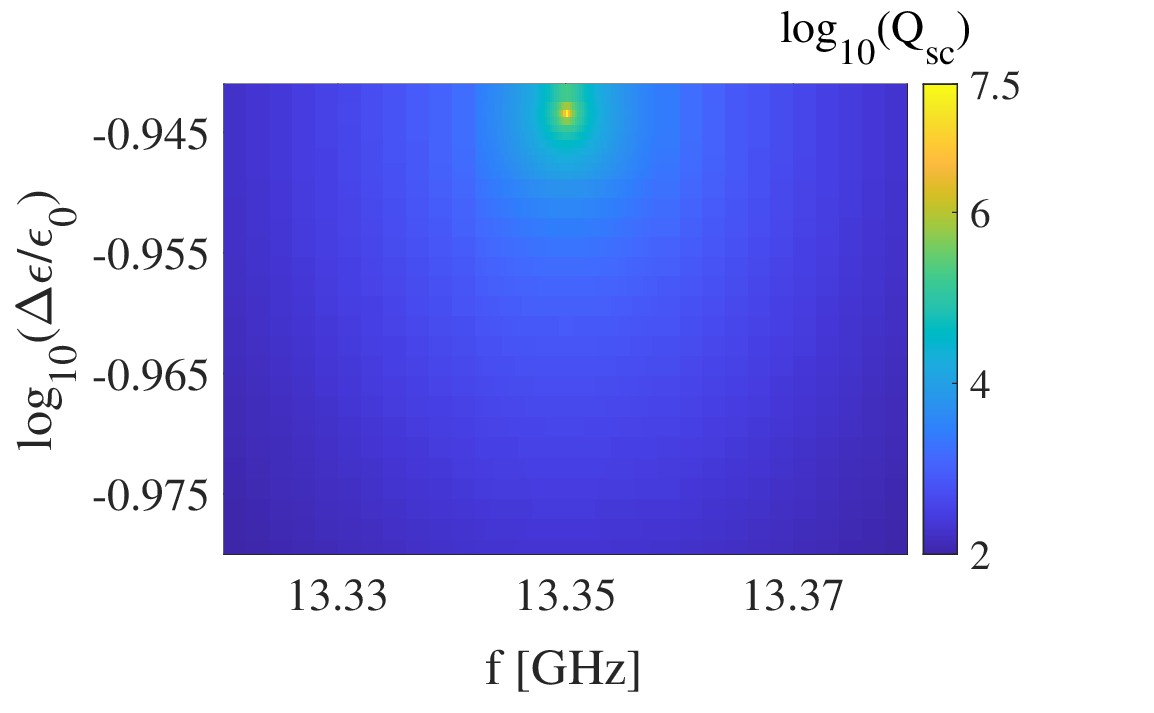}}
     \hfill
    \caption{Total scattering efficiency as defined in Eq.\,(\ref{eq:time_Qsc}), for TE mode, plot in color and in log-scale, $\textrm{log}_{10}(Q_{sc})$. Modulation frequency is $f_m=26.7$\,GHz. {Horizontal axis is incident frequency $f$, and vertical axis is $\log_{10}(\Delta\varepsilon/\varepsilon_0)$. Parametric resonance is reached for the TE mode, $l=2$, as shown by the maximum value of the $Q_{sc}$ at $f=f_m/2=13.35$\,GHz and $\Delta\varepsilon=1.366(\approx\Delta\varepsilon_{th}$).}
    }
    \label{fig:Qsc_total}
\end{figure}

To investigate further the scattering response of our system we calculate and plot in  Figure\,\ref{fig:Qsc_total} (in color)  the total scattering efficiency {(TE mode)} based on Eq.\,(\ref{eq:time_Qsc}), as we vary the frequency of incident plane wave $f$ and the modulation strength, {$\log_{10}(\Delta\varepsilon/\varepsilon_0)$. With a modulation frequency fixed at $f_m=26.7$\,GHz, for the incident frequency at half of the modulation frequency, $f=f_m/2$, the $Q_{sc}$ reaches a maximum when $\log_{10}(\Delta\varepsilon/\varepsilon_0)=-0.944$ (i.e. $\Delta\varepsilon=1.366$, very close to the $\Delta\varepsilon_{th}$ required for the TE $l=2$ mode). The calculated scattering efficiency reduces as $\Delta\varepsilon$ is further increased. As explained in \cite{sun2025formulation,canos2026revealing}, at parametric resonance the loss of a scattering system is compensated by the gain brought by the time modulation. 
Further increasing the modulation strength, in practice, gain overcomes both radiative and resistive losses (if any) and nonlinearities arise due to the strong fields, which alter the system response. Such response can not be described anymore by the model employed here, so we believe that even  the calculated scattering efficiency for $\Delta\varepsilon$ above the threshold (see  Fig.2) may not be physical.} 


\begin{figure}[t]
    \centering
    \subfloat[]{
    \includegraphics[width=0.45\textwidth]{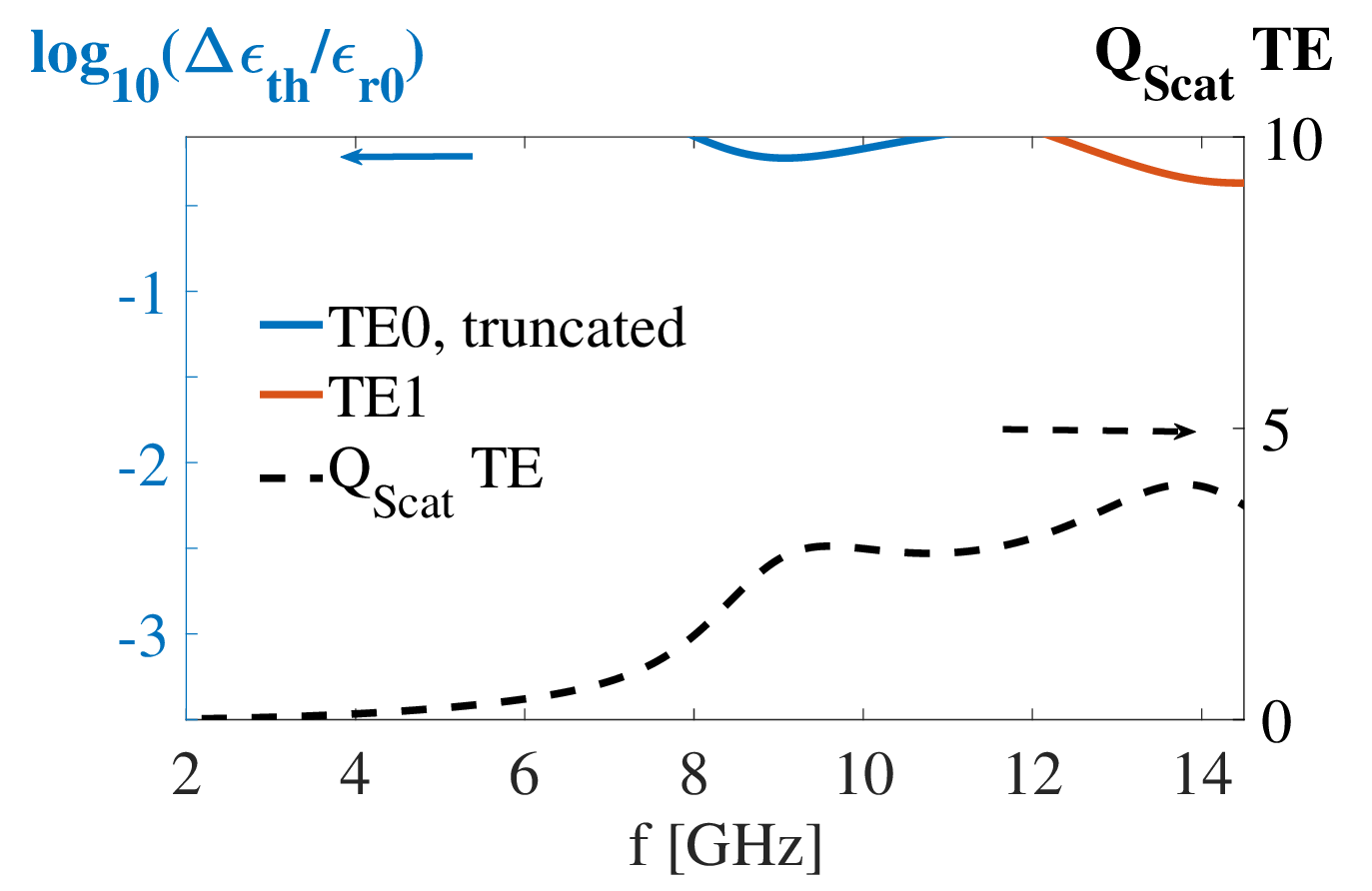}}
    \subfloat[]{
    \includegraphics[width=0.45\textwidth]{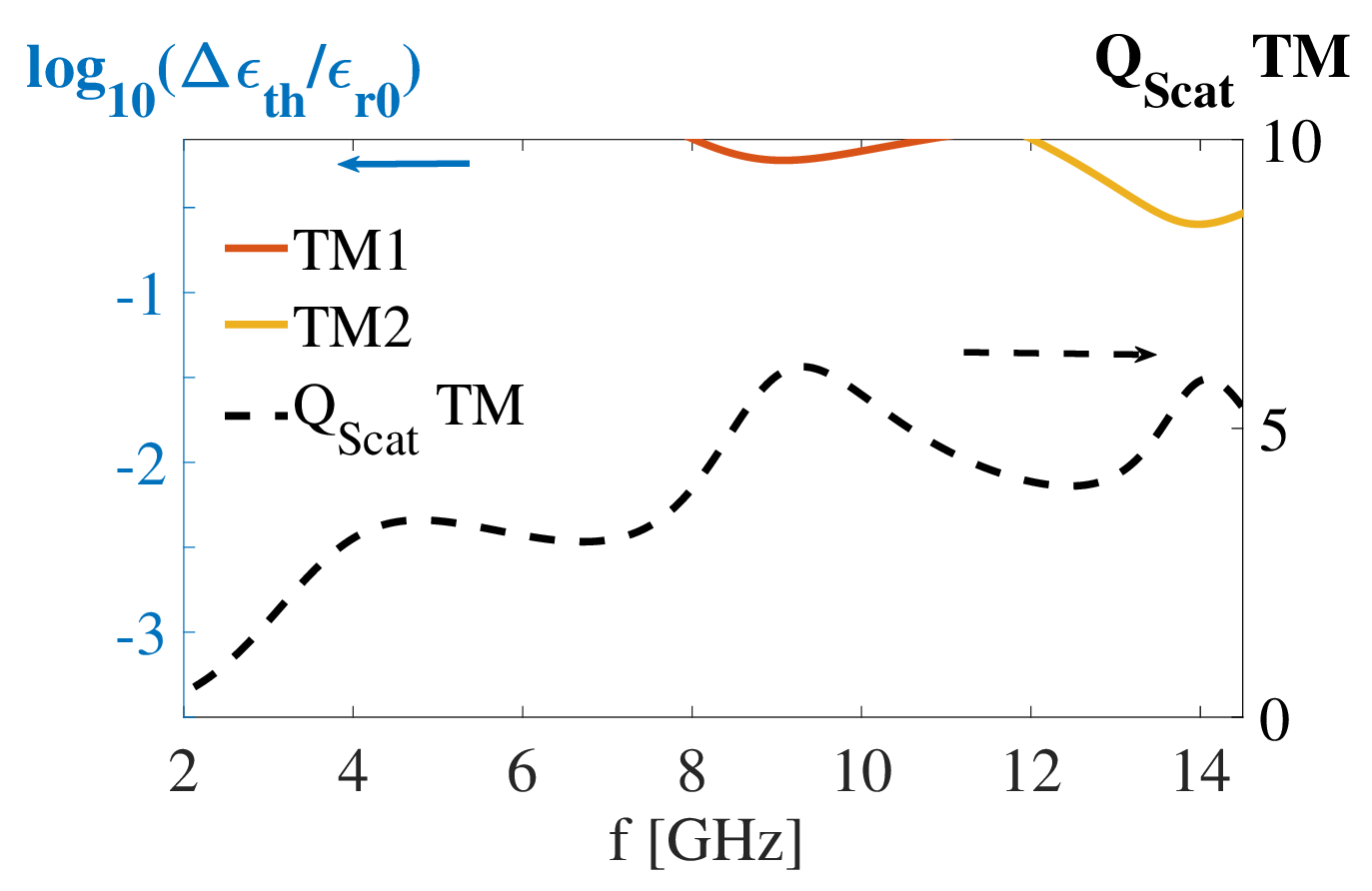}}
    \caption{Threshold values of the modulation strengths $\Delta\varepsilon_{ th}$ for parametric resonances ($\varepsilon_{r0}=6,\,R=5$\,mm, modulation frequency always satisfies $f_m=2f$, colored solid curves, left axes), and scattering efficiencies of cylinders with static relative permittivity $\varepsilon_{r}=6$ (black dashed curve, right axes). (a) is for TE polarization, (b) is for TM polarization.}
    \label{fig:eps6}
\end{figure}

Reducing the relative permittivity of a time-invariant cylinder results in broader Mie resonances, shown as the black dashed curves in Fig.\,\ref{fig:eps6}, referenced to the right axes (TE in (a) and TM in (b)). The radius of the cylinder is unchanged ($R=5$\,mm). The calculated threshold $\Delta\varepsilon$ based on the truncated eigenvalue problem for each order $l$ are plotted in different colors in Fig.\,\ref{fig:eps6}, referenced to the left axes. The higher required threshold modulation strengths compared to the case of a high permittivity cylinder are attributed to the decreased quality factor of the Mie resonances, exhibited by the time-invariant reference cylinder. {Even higher increase in the threshold modulation strengths is expected if a lossy cylinder is considered, as losses reduce even more the quality factor. This effect of losses was verified in more realistic time-varying systems, where the dispersive material properties were taken into account, and losses were included (e.g. the Drude-Lorentz model). Specifically, in \cite{blanco2026lattice} it was  shown that higher material losses lead to higher threshold modulation strengths as expected.}

\begin{figure} [b]
    \centering
    \subfloat[][]{
    \includegraphics[width=0.495\textwidth]{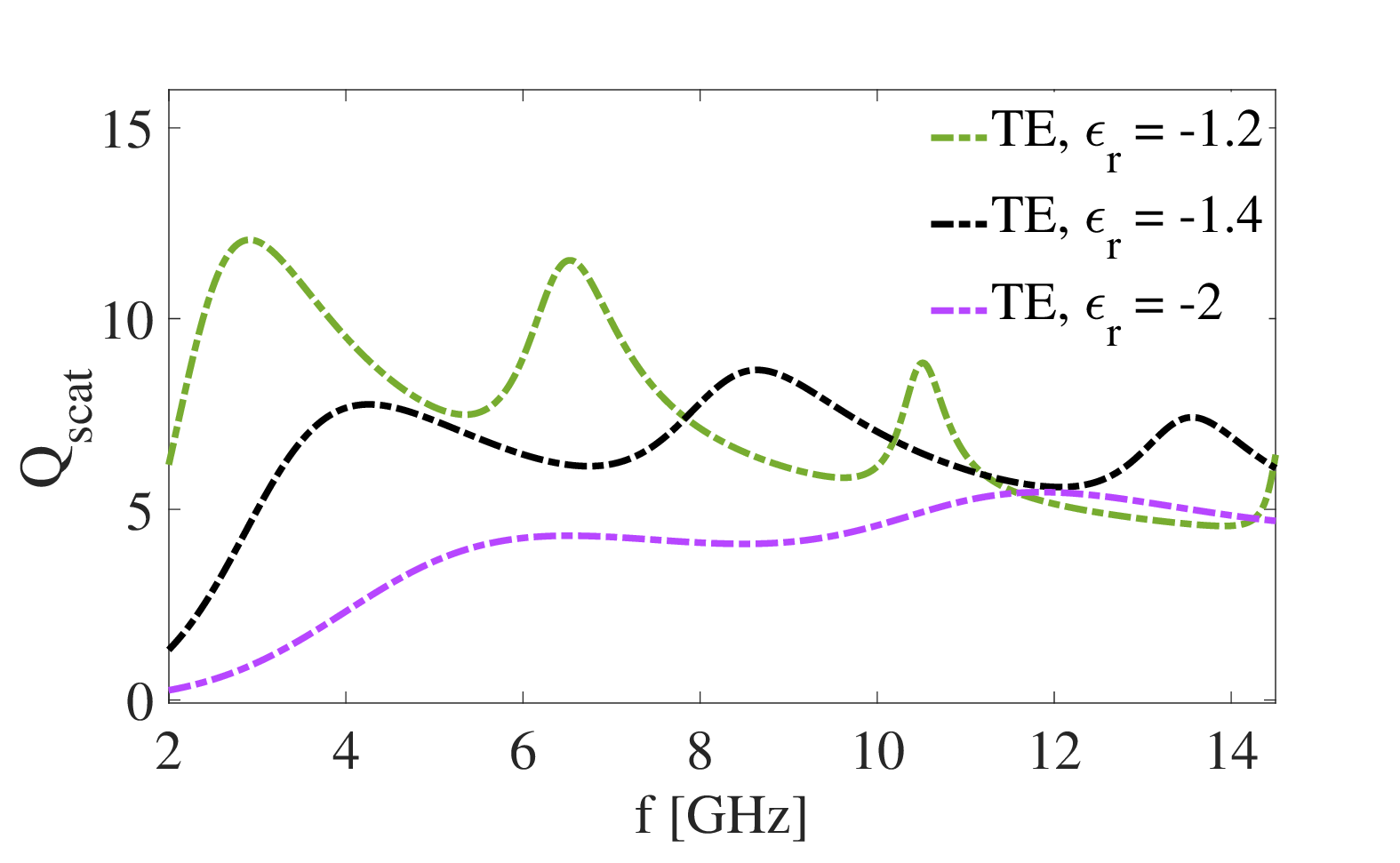}}
    \subfloat[][]{
    \includegraphics[width=0.45\textwidth]{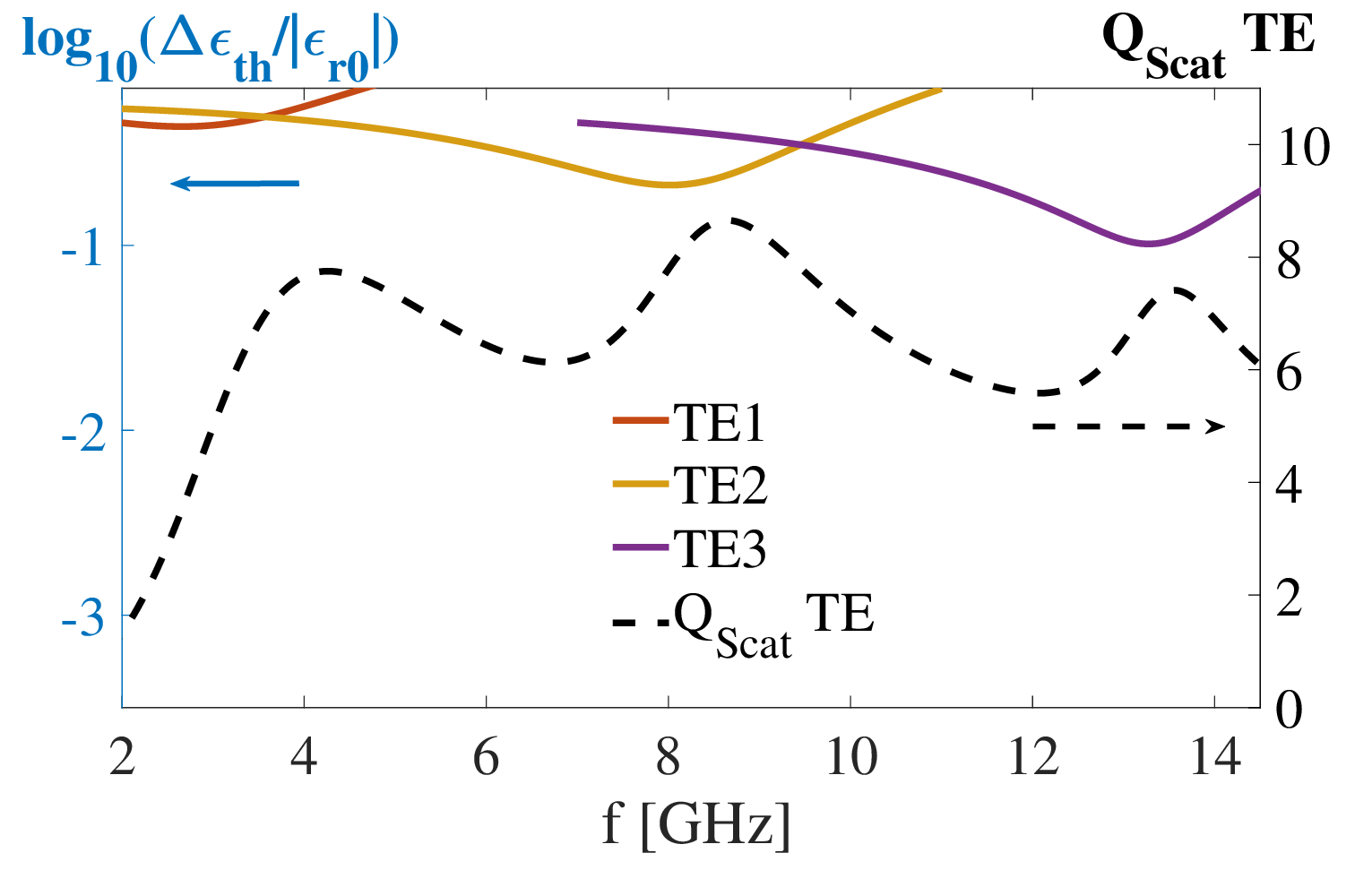}}
    \caption{(a) Scattering efficiencies of static systems  $R=5$\,mm,  $\varepsilon_{r0}=-1.2, -1.4, -2$ (free-space host medium) of TE (in different colors) and TM (blue, in different line formats) polarizations respectively. (b) Scattering efficiencies the static system, TE polarization, reference property $\varepsilon_{r0}=-1.4$, $R=5$\,mm (black dashed curve, right axes), and $\Delta\varepsilon_{th}$ required for parametric resonances under time-modulation $f_m=2f$ (colored solid curves, left axes).}
    \label{fig:Nega_eps}
\end{figure}

Another interesting case to be considered is for the cylindrical scatterer to have $\varepsilon_{r0}<0$. This applies, e.g., to metallic particles or plasma discharges below the plasma frequency or photo-excited semiconducting particles (although for the accurate study of those cases the dispersion in the material permittivity should also be taken into account \cite{blanco2026lattice,verde2026optical}). Keeping the radius $R=5$mm unchanged, and host medium as free space, $\varepsilon_h=1$, the scattering efficiencies of time-invariant scatterers where $\varepsilon_{r0}=-1.2, -1.4, -2$, are shown in Fig.\,\ref{fig:Nega_eps}(a). The results for the TM polarization are in blue, showing no peaks within the frequency range of interest, and no significant variations as $\varepsilon_{r0}$ is varied. The TE polarization, on the other hand, shows sharper Mie resonances as $|\varepsilon_{r0}|\rightarrow1$. The absence of sharp resonances for TM waves can be explained based on an analogy to the plasmonic resonances exhibited by, for example, an infinitely long metallic cylinder; localized surface plasmon resonances are not supported if physical boundaries along the direction of the electric fields are lacking, as in our TM case. 
Taking a further step to consider the time-modulated system, choosing $\varepsilon_{r0}=-1.4$, results of $\Delta\varepsilon_{th}$ based on the truncated eigenvalue problem are shown in Fig.\,\ref{fig:Nega_eps}\,(b). In this case, parametric resonance is only possible for the TE polarization. What is different from the previous cases with $\varepsilon_{r0}>0$ is that the mode $l=0$ is not possible to be parametrically amplified within the frequency range due to a much broader static Mie resonance. Despite this, the minimums of $\Delta\varepsilon_{th}$ are close to the Mie resonant frequencies, and the minimum required values decrease for higher order $l$, showing similar trend to that of the $\varepsilon_{r0}>0$ case as shown in Fig.\,\ref{fig:Par_Mie}(a). 

\section{Modulating the scattered field distribution} \label{sec:Rad}
 As is well known in single-scattering theory, superposition of the scattered harmonics of different multipolar modes allows for engineering the field distributions in the far-field range, giving, e.g., enhanced forward or backward scattering (Kerker effects). 
 We will show here that the time modulation allows to tune such a superposition (i.e. of the different multipolar modes), imposing or modulating Kerker-like effecs. 

For  a time-modulated dielectric cylinder,
the electric field of the scattered wave {($E_\varphi$ component)} of frequency $f=f^{(0)}$ (under time-modulation of $f_m=2f$), in the farfield range,  can be expressed in terms of the Mie scattering coefficients, as
\begin{equation} \label{eq:En0}
   \lim_{r \rightarrow \infty}E_{sc}^{(0)}(r,\varphi)=\frac{e^{iq_{h0}r}}{\sqrt{r}}\frac{1}{\sqrt{\pi q_{h0}}}(1-i)\left[\sum_{l=0}^{\infty}\frac{2}{1+\delta_{l0}}a_{l+}^{(0)}\cos(l\varphi)\right],
\end{equation}
\noindent where the above expression is for the TE mode, and is in the same form as the expressions in [9] for a static, unmodulated infinite cylinder. With $f_m=2f$, we have $f^{(1)}=f-f_m=-f=-f^{(0)}$. We consider $n=0,1$ as contributing harmonics when evaluating the scattering at the fundamental frequency $f=f^{(0)}$.
The electric field of the scattered wave corresponding to the negative-frequency harmonic $n=1$, is obtained in a similar way by taking the corresponding asymptotic forms of the Hankel functions, given that $q_{h1}=-q_{h0}$:
\begin{equation} \label{eq:En1}
\begin{split}
     \lim_{r \rightarrow \infty}E_{sc}^{(1)}(r,\varphi) &=  -\lim_{r \rightarrow \infty}i\sum_{l=0}^\infty \frac{2}{1+\delta_{l0}}a_{l+}^{(1)}\frac{\partial H_l(q_{h1}r)}{\partial (q_{h1}r)}i^l\cos(l\varphi)\\
     &= \frac{e^{iq_{h1}r}}{\sqrt{r}}\frac{1}{\sqrt{\pi \vert q_{h1}\vert}}(-1-i)\left[\sum_{l=0}^{\infty}\frac{2}{1+\delta_{l0}}a_{l+}^{(1)}\cos(l\varphi)\right],\\
     {\rm following}&\,\,{\rm from\,\,} \lim_{r\rightarrow\infty}\frac{\partial H_l(q_{h1}r)}{\partial (q_{h1}r)} = 
     (-i)^l\sqrt{\frac{2}{\pi|q_{h1}|r}}\,e^{iq_{h1}r}e^{-i\pi/4},
  \end{split}   
\end{equation}
\noindent where the asterisk  represents the complex conjugate. The overall phasor electric field, evaluated at $f$, is then $E_f(r,\varphi) = E_{sc}^{(0)}(r,\varphi)+\left[E_{sc}^{(1)}(r,\varphi)\right]^*$ - for details see Supplementary E. The radiation pattern then can be obtained via calculating $|E_f(r,\varphi)|^2$.

An additional degree of freedom in the design arise from the modulation phase $\beta$. As mentioned in Section\,\ref{sec:ParamMie}, by solving Eqs. (\ref{eq:Eigen_1}) and (\ref{eq:eigen}) (with $N=1$ as a simplified example), it can be concluded that only the elements $v_p(n\neq0)$ of the wave vector gain phase shifts. A more general conclusion is that the modulation phase induces a phase shift of $n\beta$ to the $n$-th scattered harmonic, as shown by the transfer matrix method in \cite{salary2018electrically}.

Following the analytical method in Section\,\ref{sec:Form}, it is clear that $v_p(n)$, and through them the phase, $\beta$,  affect the coefficients $\alpha_{l+}^{(n)}$ that determine, among others, the far-field scattering amplitudes. For the scattering at the fundamental frequency $f=f_0$ (under the condition $f_m=2f$), we have mainly contribution from  the scattered harmonics $n=0,1$. The corresponding phase shift induced by the time-modulation is then carried over as $\alpha_{l+}^{(n=0)}=(\alpha_{l+}^{(n=0)}|_{\beta=0}),\,\alpha_{l+}^{(n=1)}=e^{i\beta}(\alpha_{l+}^{(n=1)}|_{\beta=0}).$ This can be used to tune the phase difference between $a_{l+}^{(0)}$ and $a_{l+}^{(1)}$ for a certain multipolar order $l$. The overall radiation pattern may thus be controlled via varying the interference between those coefficients, as shown in the specific numerical examples discussed below.

\begin{figure}
    \centering
    \subfloat[][]{
    \includegraphics[height=0.3105\textwidth]{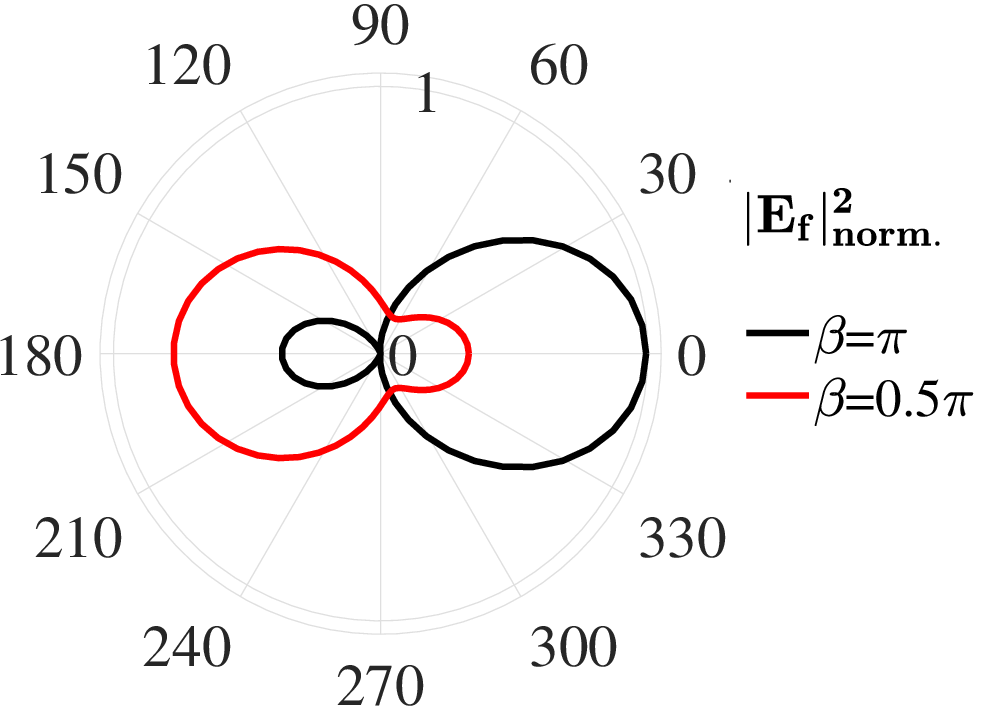}} 
    \subfloat[][]{\includegraphics[height=0.32\textwidth]{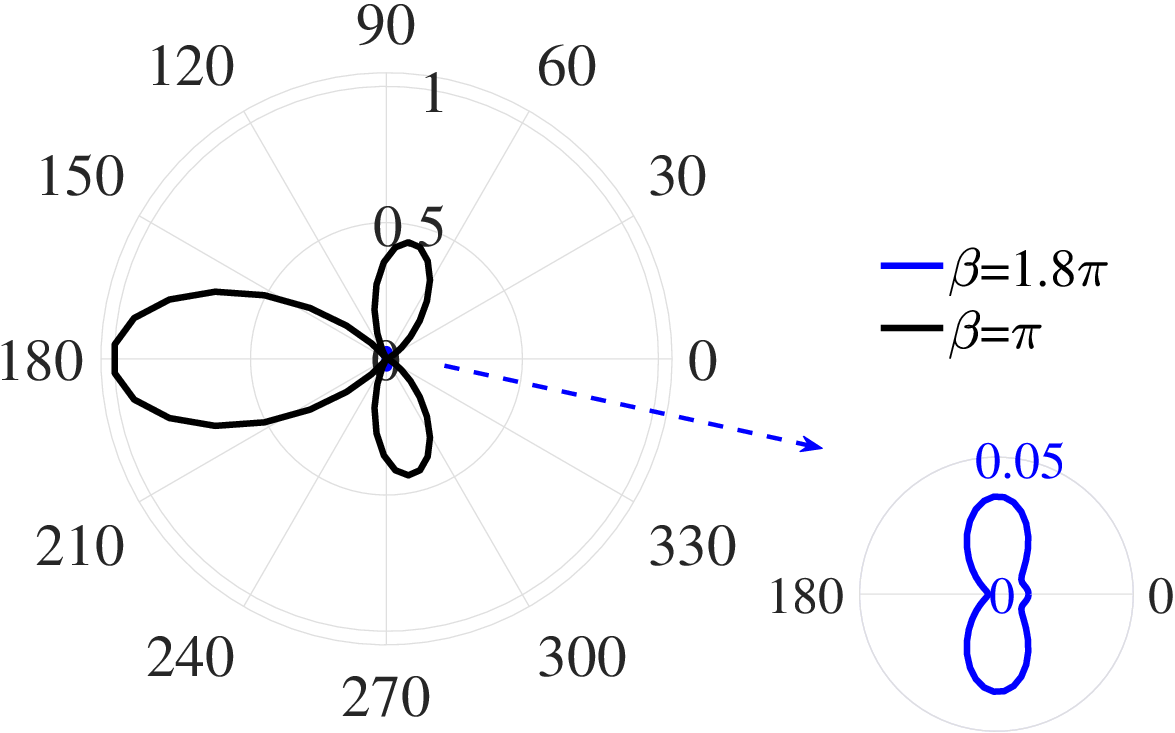}}
    \caption{Field distribution $|E_f|^2$, in the plane normal to the cylindrical axis, for a cylinder of periodically time modulated permittivity and   $\varepsilon_{r0}=12$, $R=5$mm. (a) The field is evaluated at $f=9.25$\,GHz (under $f_m=2 f$), for $\Delta\varepsilon=8.2$, $N=3$. Two cases shown: {$\beta=0.5\pi$ in red, and $\beta=\pi$} in black. The field is normalized to the maximum value for the $\beta= \pi$ case. (b) The field is evaluated at $f=12.4\,$GHz, for $\Delta\varepsilon=5.5$, $N=3$. {$\beta=\pi$} in black, and {$\beta=1.8\pi$} in blue, with a zoom-in inset showing the distribution on a much smaller scale. Both are normalized to the maximum of the $\beta=\pi$ case.}
    \label{fig:switch_Rad}
\end{figure}

An example where we excite both the zeroth and first order ($l=0,1$) TE modes, under modulation conditions for $\Delta\varepsilon$ close to but below those for the parametric resonances, 
is shown in Fig.\,\ref{fig:switch_Rad}(a). The calculation for the scattering coefficients is carried out with $N=3$ (as required for high $\Delta\varepsilon$, shown in Fig.\,\ref{fig:Par_Mie}(b)). The incident frequency is $f=9.25$\,GHz (under modulation frequency $f_m=2f=18.5$\,GHz), and the modulation strength $\Delta\varepsilon=8.2$, which is below the threshold modulation strength corresponding to the intersection of the blue and red dots in Fig.\,\ref{fig:Par_Mie}(a) (close to 9.25\,GHz). The radiated power intensity is plotted as $\vert E_f\vert ^2$, evaluated at $f^{(0)}=f$. The field distribution at this frequency contains contributions from the two scattered Fourier harmonics $n=0$ and $n=1$.  If the phase of the applied time-modulation is $\beta=0.5\pi$\,(red), predominantly backward radiation is obtained from the calculations, while $\beta=\pi$\,(black) gives predominantly forward radiation. The field distribution is normalized by the maximum of the field at $\beta=\pi$. The induced phase difference allows for the switching of the radiation main lobe at the same evaluation frequency $f=f_m/2$, through varying the phase $\beta$ of the time-modulation.  Distinct from the previous work for spherical particles\,\cite{asadchy2022parametric}, here one can switch between two contrasting regimes very easily without redesigning the scattering particle.

If, instead, the electric dipole and quadrupole moments ($l=1,2$) are excited simultaneously, at Floquet frequency $f=12.4$\,GHz ($f_m=24.8$\,GHz), and $\Delta\varepsilon=5.5$, the resultant farfield distribution is shown in Fig.\,\ref{fig:switch_Rad}(b). When $\beta=\pi$ (black), the main lobe is in the backward direction. The conditions for suppressing forward scattering (or backward scattering, in the example of Fig.\,\ref{fig:switch_Rad}(a)) are discussed in the Supplementary D).
For $\beta=1.8\pi$, the field distribution is not visibly clear when plotted on the same scale (blue), but its  strength (normalized to the maximum of the field at $\beta=\pi$) is reduced by around 20 times as shown in the inset. At $\beta=1.8\pi$, there is destructive interference between the two harmonics, resulting in the cancellation of the fields. Here the induced phase due to the time-modulation allows for `switching off\,' the far-field.

{The Kerker-like condition in the cases of Fig. 5 is offered by the interference between the adjacent multipolar modes ($l=0,1$ in Fig.5(a) and $l=1,2$ in Fig.5(b)), as well as the harmonic modes (that have the same frequency magnitude). Conditions regarding forward or backward scattering in these cases are discussed in the Supplementary E. In general, properly selecting the modulation strength and phase allows engineering of the scattering pattern, which can be of high value in the case of time-modulated-cylinder-based metasurfaces.}
Although we consider here lower-order modes that lead to relatively high modulation strengths required for exciting different modes simultaneously, one will be able to significantly decrease the required modulation strengths to practical values by switching to higher order modes. This is not considered in the current work as our purpose is to explore and demonstrate theoretically the role of the modulation phase.

\section{Conclusions}
We have developed an analytical framework combining Floquet and Mie theories for evaluating the scattering by an infinitely long cylindrical scatterer with periodically time-modulated, non-dispersive permittivity. 
Applying this framework, we identified, among others, the threshold modulation strengths required to trigger parametric Mie resonances, demonstrating that the required modulation strength is directly related to the quality factor of the corresponding Mie resonances in the static system. Our analysis further reveals that, beyond scattering enhancement, the far-field radiation pattern in the presence of multiple multipolar modes can be dynamically controlled through the modulation phase. In particular, we demonstrated the possibility of either dynamically switching the directions of the main radiation lobes or completely suppressing the far-field radiation, solely by varying the modulation phase.

Our study constitutes a comprehensive investigation of time-modulated cylindrical scatterers, highlighting their strong potential for the realization of time-modulated metamaterials and metasurfaces. Furthermore, it underscores the critical role of the modulation phase in multimode time-periodic systems, as the modulation phase governs the interference among the different modes and thus enables complete reconfiguration of the scattered-field profile.

Possible future directions include extending the analysis to cylinders with dispersive time-modulated permittivity, a configuration that could be experimentally realized, for example, using plasma discharge tubes \cite{conrads2000plasma}, where the plasma frequency can be periodically tuned in time through external excitation sources.

\begin{acknowledgments}
This work was supported by the European Union under the call ``EIC Pathfinder Open 2022" (Project PULSE, project no. 101099313). 

\end{acknowledgments}

\section*{Supplementary}
\subsection*{A: Fields in the scattering system of a
time-modulated cylinder}

We consider plane wave scattering by an infinite cylindrical scatterer made of a time-modulated medium, embedded in a homogeneous host material (of permittivity $\varepsilon_h$ and permeability $\mu_h$). The cylinder is assumed  parallel to the $z$-direction and the incident wave propagates along $x$-direction. We will express the fields in the system in cylindrical coordinates ($r,\varphi, z$), and in terms of the cylindrical vector harmonics $\mathbf{n}_l$ and $\mathbf{m}_l$. The harmonics $\mathbf{n}_l$ and $\mathbf{m}_l$ can be expressed (in terms of Bessel functions of first kind, $J$) as 

\begin{equation}
\textbf{n}_l=kJ_l(kr)\cos(l\varphi)\hat{\textbf{z}},\,\, \, \textbf{m}_l=\frac{l}{r}J_l(kr)\sin(l\varphi)\hat{\textbf{r}}-k\frac{\partial J_l(kr)}{\partial(kr)}\cos(l\varphi)\hat{\boldsymbol{\varphi}},\label{eq:mn}
\end{equation}
with $k$ the wavenumber in the medium.

For the case of TE wave scattering (i.e. electric field perpendicular to the cylinder axis), the incident electric field is along the $y$-direction, and is considered to be  
\begin{equation}
    \textbf{E}=e^{-i\omega t}e^{i\textbf{k}\cdot\textbf{x}}\hat{\textbf{y}}=\textbf{E}'e^{-i\omega t}, \label{eq:E1}
\end{equation}
where $\textbf{E}'$ is not dependent on time.
The corresponding magnetic field can be obtained from Faraday's law, $  \nabla\times\textbf{E}=i\omega\mu_0\mu_h\textbf{H}$ (for harmonic waves of the form $e^{-i\omega t}$), as 
\begin{equation}
    \textbf{H}=\textbf{H}'e^{-i\omega t},\, \, \, \textbf{H}'=\frac{1}{Z_0}\frac{ck}{\omega\mu_h}e^{i\textbf{k}\cdot\textbf{x}}\hat{\textbf{z}}=\frac{1}{Z_h}e^{i\textbf{k}\cdot\textbf{x}}\hat{\textbf{z}},
\end{equation}
\noindent with $Z_0=\sqrt{\mu_0/\varepsilon_0}$ being the wave impedance in free space,  $k=k_0=\omega\sqrt{\varepsilon_h \mu_h}/c$ is the wavenumber in the host material and  $Z_h=Z_0\sqrt{\mu_h/\varepsilon_h}$ is the wave impedance in the host material.

In expressing the fields in cylindrical waves we  use the useful expansion \cite{ii1941electromagnetic}
\begin{equation}
e^{i\textbf{k}\cdot\textbf{x}}=e^{ikr\cos\varphi}=\sum_{l=0}^{\infty} \frac{2}{1+\delta_{l0}}i^l J_l(kr)\cos(l\varphi).
\end{equation}
From the above equation one can immediately see that the incident magnetic field  $\textbf{H}'$ can be expanded as
\begin{equation}
    \textbf{H}'(\textbf{r})=\frac{1}{Z_h}\sum_{l=0}^{\infty} \frac{2}{1+\delta_{l0}}i^l J_l(kr)\cos(l\varphi)\,\hat{\textbf{z}}= \sum_{l=0}^{\infty}\frac{2}{1+\delta_{l0}}\frac{i^l}{k} \frac{1}{Z_h} \textbf{n}_l.
    \label{eq:H1}
\end{equation}

In expressing the electric field in terms of the vector harmonics one can use Ampere's law for harmonic waves, i.e. 
 $-i\omega\varepsilon_0\varepsilon_h\textbf{E}'(\textbf{r}) = \nabla\times\textbf{H}'(\textbf{r})$.
 Using cylindrical coordinates in the curl evaluation and the above form of ${\bf H}'$, one can obtain
\begin{align}
    \textbf{E}'(\textbf{r}) = \frac{i}{\omega\varepsilon_0\varepsilon_h}\frac{1}{r}\frac{\partial H_z}{\partial \varphi}\hat{\textbf{r}}-\frac{i}{\omega\varepsilon_0\varepsilon_h}\frac{\partial H_z}{\partial r}\hat{\boldsymbol{\varphi}}
    =\frac{i}{k}\sum_l \frac{2}{1+\delta_{l0}}i^l [\frac{l}{r}J_l(kr)\sin(l\varphi)\mathbf{\hat{r}}- k \frac{\partial J_l(kr)}{\partial(kr)}
    \cos(l\varphi)\mathbf{\hat{\varphi}}]. 
\end{align}
Thus, 
 \begin{equation}
  \textbf{E}'(\textbf{r}) = \textbf{E}_\mathrm{inc}(\textbf{r})=i\sum_l\frac{2}{1+\delta_{l0}}\frac{i^l}{k}\textbf{m}_l=i\sum_l a_l^0\textbf{m}_l, \hspace{4mm}  a_l^0= \frac{2}{1+\delta_{l0}}\frac{i^l}{k}.
  \label{eq:E1}
\end{equation}
with $a_l^0$ the incident wave expansion coefficients of Eq. (5) of the main text.

For the corresponding magnetic field,    \begin{equation}
\textbf{H}'(\textbf{r})=\sum_{l=0}^{\infty} {b_l^0}' \textbf{n}_l,  \hspace{4mm} {b_l^0}'=\frac{2}{1+\delta_{l0}}\frac{i^l}{k} \frac{1}{Z_h}, 
\end{equation}
giving $a_l^0/{b_l^0}'=Z_h$.

Given the orthogonality of $\mathbf{m}$ and $\mathbf{n}$ vector functions and the cylindrical  symmetry of the problem (due to the cylindrical scatterer), an incident wave of the form of Eq. (\ref{eq:E1}) propagating in the plane perpendicular to the cylinder  will excite waves inside the cylinder and scattered waves having the same multipolar functional form, i.e. the electric fields will be linear combinations of functions $\mathbf{m}_l$, as given in Eqs. (6) and (7) of the main text.


Similarly to the incident wave coefficients, for the scattered harmonics with Floquet order $n$, 
\begin{align}
    \textbf{H}_{h+}^{(n)}(\textbf{r})&=\sum_l b_{l+}^{(n)'}\textbf{n}'_l(q_{hn},r),\\
    \textbf{E}_{h+}^{(n)}(\textbf{r})&=i\sum_l a_{l+}^{(n)'}\textbf{m}'_l(q_{hn},r),
\end{align}
and one can show that $a_{l+}^{(n)'}/b_{l+}^{(n)'}=Z_h$, as the scattered waves also propagate in the host medium.

For the fields inside the time-modulated cylinder, expressed as in Eqs. (6) and (7) of the main text, the different Fourier components will have the form 
\begin{align} 
  \textbf{E}_c^{(n)}(\textbf{r}) &=i\sum_{p=-N}^N v_{p}(n)\sum_l a'_{l;p}\textbf{m}_l(q_{p},\textbf{r}), \\
  \textbf{H}_c^{(n)}(\textbf{r}) &=\sum_{p=-N}^N v_{p}(n)\sum_l b'_{l;p}\textbf{n}_l(q_{p}.\textbf{r}).\label{eq:H_cyl}
\end{align}

It also holds that
\begin{equation}
   \nabla\times\textbf{m}_l(q_{p},r)=q_p\textbf{n}_l(q_{p},r).\label{eq:mnrel}
\end{equation}
For each $p,n,l$, using the Faraday's law and Eq.\,(\ref{eq:mnrel}), we arrive at
\begin{equation}
\begin{split}
    a_{l;p}'\nabla\times\textbf{m}_l(q_{p},r)&=-b_{l;p}'\,\mu_0\mu_r(n\omega_m-\omega)\textbf{n}_l(q_{p},r),\\
    q_p a_{l;p}'&=Z_0\frac{\mu_r(\omega-n\omega_m)}{c}b_{l;p}'\,.
\end{split}
\end{equation}

To summarize, we have the following relations between the scattering coefficients: $a^{(n)}_{l+}=a^{(n)'}_{l+}/a_l^0;$ $\,q_p a_{l;p}=b_{l;p}\mu_r2\pi(f-nf_m)/c$ , $a_l^0 = \sqrt{\mu_h/\varepsilon_h }\,b_l^0$ and $a^{(n)}_{l+}= \sqrt{\mu_h/\varepsilon_h }\,b^{(n)}_{l+}$ as stated in the main text.

For the TM polarization, the incident magnetic field is in the $y$ direction. The incident electric field, expressed in a similar way to Eq. (S3), is
\begin{equation}
  \textbf{E}=\textbf{E}'(\textbf{r})e^{-i\omega t},\hspace{2mm} \textbf{E}'(\textbf{r})=e^{i\mathbf{k}\cdot\mathbf{x}}\hat{\textbf{z}},
  \end{equation}
 with cylindrical expansion 
 \begin{equation}
 \textbf{E}'(\textbf{r})=\sum_l\frac{2}{1+\delta_{l0}}i^l J_l(kr)\cos(l\varphi)\,\hat{\textbf{z}}\\
  =\sum_l\frac{2}{1+\delta_{l0}}\frac{i^l}{k}\textbf{n}_l.
\end{equation}
From Faraday's law, we then have 
\begin{equation}
    \begin{split}
        \textbf{H}'(\textbf{r})=-i\sum_l\frac{2}{1+\delta_{l0}}\frac{i^l}{k}\frac{kc}{Z_0\omega\mu_h}\textbf{m}_l.
    \end{split}
\end{equation}
\noindent These fields are very similar to the corresponding ones for TE polarization, except for: 1) the constant coefficient $i$ or $-i$ outside the summation, representing the directions of the fields; 2) the exchange of the cylindrical vector harmonics $\textbf{m}_l,\,\textbf{n}_l$ within the summation, again related to the fields directions as we changed polarization.

\subsection*{B: Truncated eigenvalue problem}
\rm Taking into account only the two dominant Fourier harmonics, $n=0,1$, Eq.\,(3) in the main text reduces to the truncated eigen-value problem below:
\begin{equation}
    \begin{pmatrix}
    (2\pi f/c)^2\varepsilon(0)  & (2\pi f/c)^2\varepsilon(-1)\\
         [2\pi (f-f_m)/c]^2\varepsilon(1) & [2\pi (f-f_m)/c]^2\varepsilon(0)
    \end{pmatrix}
    \begin{pmatrix}
        E^{(0)}\\
        E^{(1)}
    \end{pmatrix}=q^2\begin{pmatrix}
        E^{(0)}\\
        E^{(1)}
    \end{pmatrix},
\end{equation}
\noindent where the Fourier coefficients in the time-modulated relative permittivity (see Eq.\,(1) of the main text) are: $\varepsilon(0)=\varepsilon_{r0},\,\,\varepsilon(1)=e^{i\beta}\Delta\varepsilon /2,\,\,\varepsilon(-1)=e^{-i\beta}\Delta\varepsilon /2$. The eigenvalue problem is then  expressed as
\begin{equation} \label{eq:trunc_eig}
     \det \left[ \begin{pmatrix}
     (2\pi f/c)^2\varepsilon_{r0}  & (2\pi f/c)^2e^{-i\beta}\Delta\varepsilon/2\,\\
         (2\pi (f-f_m)/c)^2e^{i\beta}\Delta\varepsilon/2\, & (2\pi (f-f_m)/c)^2\varepsilon_{r0}
    \end{pmatrix} -q^2\bf{I}\right] = 0,
\end{equation}
\noindent with $\bf{I}$ being a 2$\times$2 identity matrix. 
Given that $f_m=2f$, the above equation leads to
\begin{equation} \label{eq:q}
    [q^2-(\frac{2\pi f}{c})^2(\varepsilon_{r0}-\frac{\Delta\varepsilon}{2})][q^2-(\frac{2\pi f}{c})^2(\varepsilon_{r0}+\frac{\Delta\varepsilon}{2})]=0.
\end{equation}
\noindent The eigenvectors then follow by inserting the $q_{1,2}$ values from Eq.\,(\ref{eq:q}) into Eq.\,(\ref{eq:trunc_eig}), as listed in Section 3 of the main text. The normalization of the eigenvector $\textbf{v}_p$ does not affect the parametric resonance conditions, which can be shown  by writing down the determinant of the 4$\times$4 matrix of Eq.\,(13) in the main text, and equating it to zero. Any normalization factor to the eigenvectors is carried over as a constant multiplication factor in solving the equation. For a more general case, considering both positive and negative $n$ (e.g. $N=1,\,n=-1,0,1$, without truncation under the parametric resonance condition), the normalization factor to $\textbf{v}_{p}$ scales the $a_{l;p},b_{l;p}$ on the LHS of Eqs.\,(8), (9) of the main text, which are the coefficients describing the time-modulated material inside the cylinder. Mathematically, this is due to the terms with $\delta_{n0}$ on the RHS of the equations, describing a fixed magnitude of the incident waves. In terms of the physical insight, the magnitude of the $E$ field inside the cylinder does not change if we change the normalization factor affecting $v_p(n)$ (as can be seen from the expression based on Eq.\,(S11), or equivalently Eq. (6) of the main text). The normalization factor to the eigenvectors $\textbf{v}_{p}$ therefore has no physical meaning.

\subsection*{C: Threshold modulation strengths and Mie resonances}
As discussed in \cite{canos2026revealing}, parametric resonances are realized when the corresponding complex eigenfrequency of the static resonant system is brought to the real axis upon the periodic modulation. Recall that for the static cylinder, the scattering coefficients are (here it is equivalent to the time-varying model with $N=0$):
\begin{equation} \label{eq:eigenfreq}
    \begin{split}
        \textrm{TE\,\,mode:\,\,} a_{l+} &= \frac{\sqrt{\mu_h/\varepsilon_h}J_l(qR)J_l'(kR)-\sqrt{\mu_r/\varepsilon_{r0}}J_l(kR)J_l'(qR)}{\sqrt{\mu_r/\varepsilon_{r0}}H_l(kR)J_l'(qR)-\sqrt{\mu_h/\varepsilon_h}J_l(qR)H_l'(kR)},\\
        \textrm{TM\,\,mode:\,\,}b_{l+}&=\frac{\sqrt{\mu_h/\varepsilon_h}J_l'(qR)J_l(kR)-\sqrt{\mu_r/\varepsilon_{r0}}J_l'(kR)J_l(qR)}{\sqrt{\mu_r/\varepsilon_{r0}}J_l(qR)H_l'(kR)-\sqrt{\mu_h/\varepsilon_h}H_l(kR)J_l'(qR)},
    \end{split}
\end{equation}
\noindent for the TE and TM modes respectively. The complex eigenfrequencies for either mode can be evaluated via setting the denominators of the scattering coefficients to 0. 

\noindent The imaginary parts of the complex eigenfrequencies represent losses and are related to the quality factors of the resonance. It is worth noting that the TE0 and TM1 modes not only occur at the same frequency (as proved previously for the static infinite cylindrical scatterers\,\cite{mavidis2020polaritonic}), but they exhibit identical threshold values of the modulation strengths as in the two plots of Fig.\,1 in the main text. To relate this observation to the eigenfrequency of the static dielectric cylinder, conditions based on Eq.\,(\ref{eq:eigenfreq}) become (with $J_0'(qR)=-J_1(qR)$ and $q/k=\sqrt{\varepsilon_{r0}\mu_r}/\sqrt{\varepsilon_h\mu_h}$),

\begin{equation}
    \begin{split}
        {\rm TE0:}\,\,& \sqrt{\mu_r\varepsilon_h}H_0(kR)J_1(qR)-\sqrt{\mu_h\varepsilon_{r0}}J_0(qR)H_1(kR)\rightarrow0,\\
        {\rm TM1:}\,\,&\sqrt{\mu_r\varepsilon_h}J_1(qR)H_0(kR)-\sqrt{\mu_h\varepsilon_{r0}}H_1(kR)J_0(qR)\\-&[\sqrt{\mu_r\varepsilon_h}J_1(qR)H_1(kR)/(kR)-
        \sqrt{\mu_h\varepsilon_{r0}}H_1(kR)J_1(qR)/(qR)]\\
        &\rightarrow0,\\
        \textrm{where we used}\,\,J_l'(x)& =J_1(x)/x-J_2(x)\\
        &=(J_0(x)-J_2(x))/2 \,\,\textrm{,  removing the dependence on}\,\,J_2(x)
    \end{split}
\end{equation}
\noindent Here the terms within the square bracket of TM1 mode reduces to 0 only if $\mu_r=\mu_h$ (as they are assumed to be both equal to 1 in this work), and this results in the same complex eigenfrequency for the TE0 and TM1 modes, which then explains the same modulation strength required for parametric resonance. 

\begin{figure}
    \centering
    \includegraphics[width=0.6\linewidth]{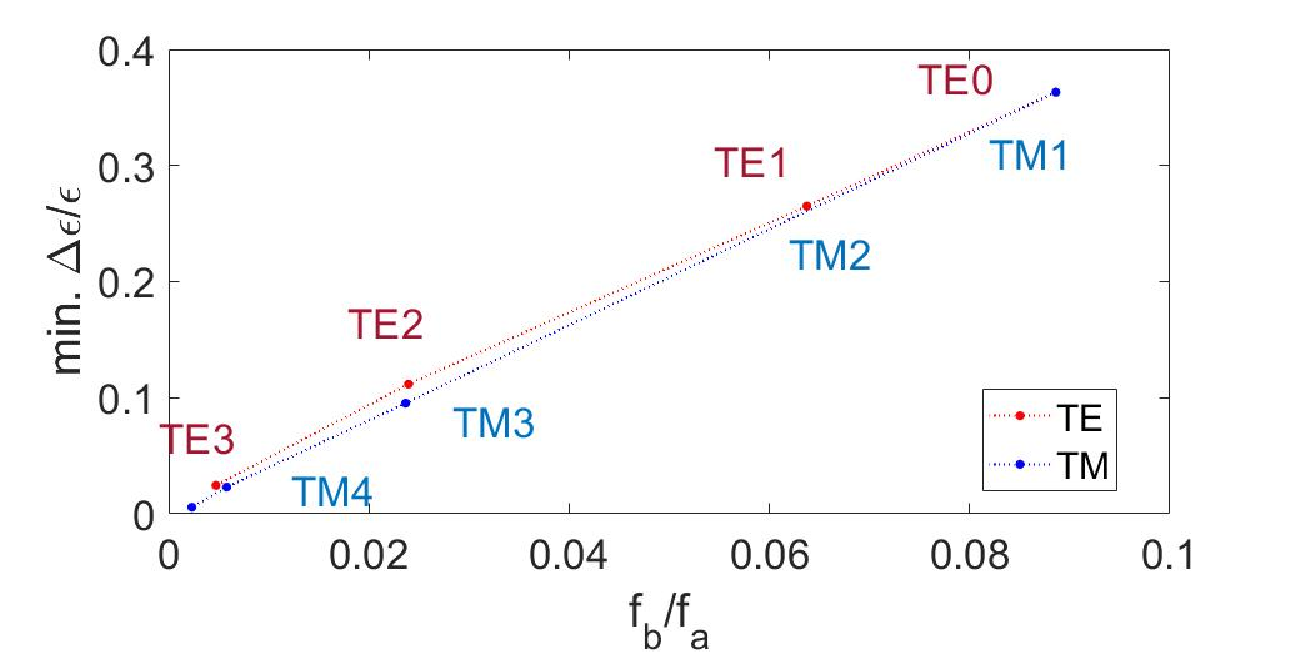}
    \caption{Minimum points of the threshold modulation strength (linear scale, values from Fig.\,1, panels (b) and (d) of the main text) against the ratios of imaginary to real parts ($f_b/f_a$) of the complex eigenfrequencies evaluated for a static cylinder with $\varepsilon_r=12$ and $R=5$\,mm.}
    \label{fig:mod_eig}
\end{figure}

Following the same approach, we can numerically search for the complex eigenfrequencies $f_{\rm eig} = f_a+if_b$ for each multipolar order $l$. Based on the Eqns.\,(20) and (30) in \cite{canos2026revealing}, where a sinusoidal modulation scheme in permittivity was also adopted, the modulation threshold is approximately proportional to the ratio of the imaginary part and real part of the eigenfrequency. Therefore, we proceed to investigate the relationship between the minimum modulation threshold (minimum values in Fig.\,1, (b) and (d), main text, of each solid curve) and the numerically calculated $f_b/f_a$ for each multipolar mode - a linear relationship is found for both TE and TM mode as shown in Figure\,S.\ref{fig:mod_eig}. The results indicate that the minimum threshold modulation strength is dependent on the losses/quality factors of the eigenmodes of the static system, with an approximate relationship $\Delta\varepsilon_{th}/\varepsilon_{r0}\approx2/Q$ where $Q=f_a/2f_b$.

\subsection*{D: Extinction and Absorption}
\begin{figure}
    \centering
    \subfloat[][]{
    \includegraphics[width=0.5\linewidth]{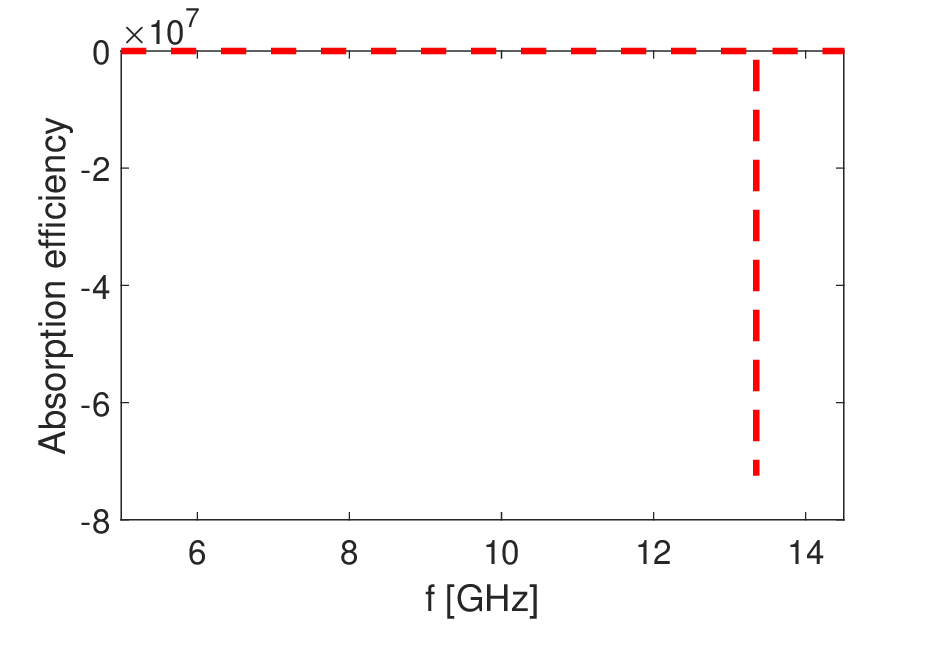}\label{fig:S2a}}
    \subfloat[][]{
    \includegraphics[width=0.4\linewidth]{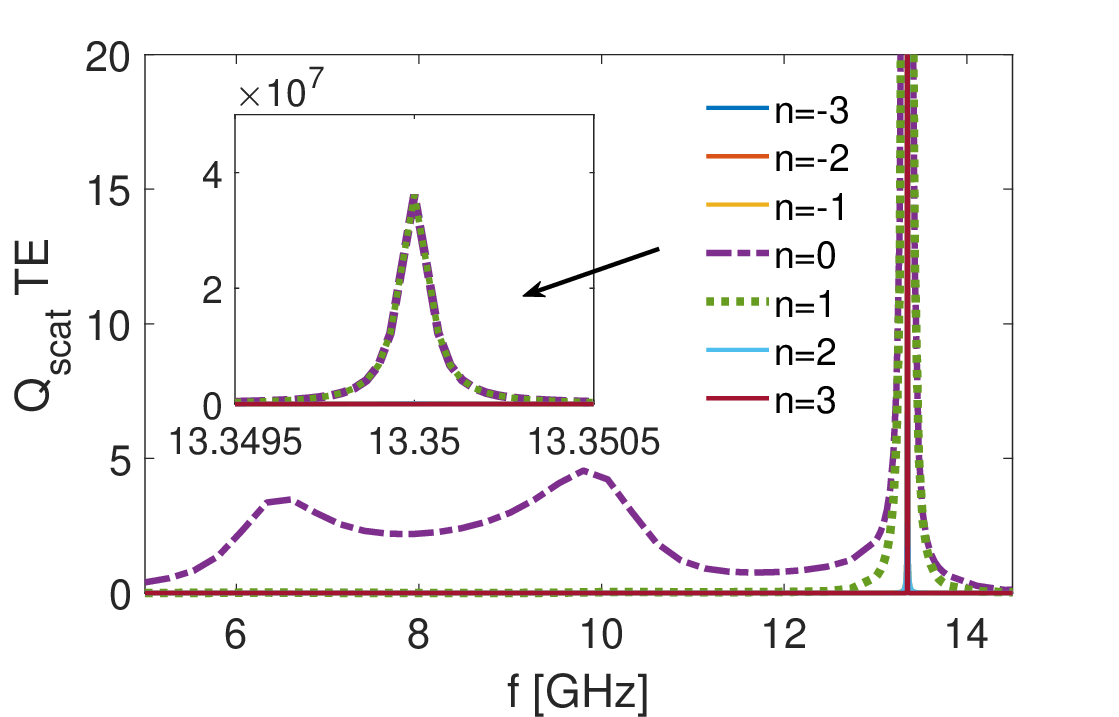}\label{fig:S2b}
    }
    \caption{{Absorption efficiency and scattering efficiency of a modulated cylinder, with parameters $\Delta\varepsilon=1.366$, which is the threshold modulation strength calculated for parametric resonance at 13.35GHz ($l=2$ resonance), and $f_m=26.7$GHz. (a) Sharp dip showing negative absorption at 13.35GHz. (b) Scattering efficiency for the different frequency harmonics plotted over incident frequency $f$. The dominant frequency harmonics are $n=0,1$ as shown in the inset, which has a different scale of the axes to show the resonant features.}}
    \label{fig:placeholder}
\end{figure}

{The total extinction efficiency of the time-modulated cylinder can be calculated as  
\begin{equation}
    Q_{ext} = -\frac{2}{\vert q_{h0}\vert R}\sum_l \frac{2}{1+\delta_{l0}}Re\{a_{l+}^{(0)}\},
\end{equation}
where extinction represents the interference between the incident wave (at $f$) and the scattered field, where only the $n=0$ frequency harmonic (of  $f^{(0)}=f$) has non-vanishing contribution.
The absorption efficiency then follows as $Q_{abs}=Q_{ext}-Q_{sc}$. The calculated absorption efficiency for TE mode, $f_m=26.7$GHz , $\Delta\varepsilon=1.366$, is plotted in Fig.   S\ref{fig:S2a}, showing that the values are close to zero within a wide frequency range except for $f=13.35$ GHz, where the absorption is negative with a large magnitude. This indicates that at conditions close to parametric resonance, power is gained in the time-modulated system through the external modulation. This gain is shown also in the scattering efficiency, which is highly amplified, as shown in Fig. S\ref{fig:S2b}. The legends there indicate the frequency harmonics considered (corresponding to frequencies $f^{(n)}=f-nf_m$). The fundamental harmonic is $n=0$, which shows the resonant features of the static system, apart from the frequency regime close to 13.35 GHz, where both the $n=0,1$ harmonics are amplified significantly.
The inset shows that the $n=0,1$ frequency harmonics are the dominant ones under parametric resonance. It should be noted that the horizontal axes in Fig. S2 are incident (evaluation frequency), but the scattering efficiencies $Q^{(n)}$ if plotted over $f-nf_m$ will appear as distinct peaks at the corresponding frequencies (positive or negative).}
\subsection*{E: Suppressing scattering in forward or backward directions}

In order to calculate the scattered far-field evaluated at $f=f_0$ taking into account the contributions from the fields of both positive and negative frequencies ($n=0,1$ respectively), we express the total physical scattered $E-$field (its $\varphi$ component) as below:
\begin{equation}
\begin{split}
    \Re\{E_{sc}(t)\}&=\Re\{E^{(0)}e^{-i2\pi f t}+E^{(1)}e^{i2\pi f t}\}\\
    &=\Re\{[E^{(0)}+E^{(1)*}]e^{-i2\pi f t}\},
\end{split}
\end{equation}
\noindent where $E^{(0)}=E_{sc}^{(0)}(r,\varphi),E^{(1)}=E_{sc}^{(1)}(r,\varphi)$ follow the expressions in Eq. (15) and Eq. (16) of the main text. 
\noindent The total phasor field at $f=f_0$ can be expressed as $E_f(r,\varphi)=E^{(0)}+E^{(1)*}$.

\begin{figure}
    \centering
    \includegraphics[width=0.3\linewidth]{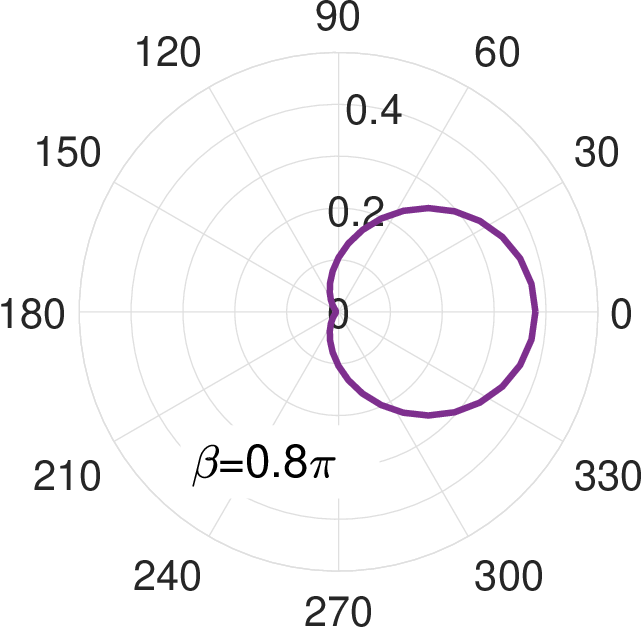}
    \caption{{Field distribution $|E|^2$, with parameters as in the case of Fig. 5(a) of the main text ($f=9.25$\,GHz (under $f_m=2 f$), for $\Delta\varepsilon=8.2$, $N=3$), except for $\beta=0.8\pi$, as predicted by the conditions in Eq.(S25) above.  The field is normalized to the maximum value for the $\beta=\pi$ case.}}
    \label{fig:S3}
\end{figure}

For the case shown in Fig.\,5(a) of the main text, the main contributing multipolar TE modes are of $l=0,\,1$, for both the Fourier harmonics $n=0,1$. 
{Based on Eq. (15) and (16), writing $A_{0,1}=a_{l=0,1,+}^{(0)}\vert_{\beta=0},B_{0,1}=a_{l=0,1,+}^{(1)*}\vert_{\beta=0}$ (all with the benchmark values at $\beta=0$), the condition for zero forward or backward scattering intensity  can be expressed as below:
\begin{equation}
    \begin{split}
        &\vert A_0+2A_1\cos(\varphi)\vert=\vert B_0+2B_1\cos(\varphi)\vert \\
        &\beta = -\textrm{Arg}\left[A_0+2A_1\cos(\varphi)\right]+\textrm{Arg}\left[B_0+2B_1\cos(\varphi)\right],
    \end{split}
\end{equation}
\noindent where for $\varphi=0$ we obtain the modulation phase required for zero forward scattering, and for $\varphi=\pi$ the modulation phase giving zero backward scattering.
The calculated coefficients under $f=9.25$GHz, $f_m=2f,\,\Delta\varepsilon=8.2$, as in Fig.5(a) of the main text, combined with the above equations, give  the modulation phase that totally suppresses the backwards scattering to be $\beta\approx0.8\pi$. The corresponding farfield is shown  in Fig. S\ref{fig:S3}. The back-scattering is suppressed in this case (unlike the case of Fig. 5(a)), although with a broader main lobe. Under this set of modulation parameters, the requirement regarding the amplitudes of coefficient is not met for the case of zero forward scattering. In order to suppress the forward scattering fully, careful selection of the modulation parameters is required. Note that the $a_{l+}^{(n)}$ in the above equations represent coefficients solved at $\beta=0$, and that the modulation phase $\beta$ only leads to phase shifts of the coefficients of $n=1$. 

For the second case considered in the main text, shown  in Fig.\,5(b), at 12.4\,GHz, the simultaneously excited multipolar orders are of $l=1,\,2$, again for both the $n=0,1$ Fourier harmonics. The corresponding conditions for zero forward or backward scattering are of a similar form as in Eq.(S25) above, i.e.:
\begin{equation}
\begin{split}
 &\vert A_1\cos(\varphi)+A_2\cos(2\varphi)\vert = \vert B_1\cos(\varphi)+B_2\cos(2\varphi)\vert\\
    &\beta = -\textrm{Arg}\left[A_1\cos(\varphi)+A_2\cos(2\varphi)\right]+\textrm{Arg}\left[B_1\cos(\varphi)+B_2\cos(2\varphi)\right].
\end{split}
\end{equation}
}
Based on the above expressions, we were able to tune the modulation phase parameter, i.e. $\beta$, to minimize or maximize the scattering intensity in certain directions. 

\end{document}